\documentclass[10pt, conference, letterpaper]{IEEEtran}
\usepackage[pass]{geometry}
\usepackage{nopageno}
\usepackage{etex}
\usepackage{subcaption}
\usepackage{multirow}
\usepackage[font=footnotesize]{caption}
\usepackage{algorithm}
\usepackage{graphicx}
\usepackage{algpseudocode}
\usepackage{amsfonts}
\usepackage{amsthm}
\usepackage{amsmath}
\usepackage{amssymb}
\usepackage{array}	
\usepackage{bbm}
\usepackage{cite}
\usepackage{dsfont}
\usepackage{epstopdf}
\usepackage{float}
\usepackage[T1]{fontenc}
\usepackage{color,soul}

\usepackage{algorithmicx}
\usepackage{algpseudocode}

\usepackage{comment}

\newcolumntype{L}[1]{>{\raggedright\let\newline\\\arraybackslash\hspace{0pt}}m{#1}}
\newcolumntype{C}[1]{>{\centering\let\newline\\\arraybackslash\hspace{0pt}}m{#1}}
	\newcolumntype{R}[1]{>{\raggedleft\let\newline\\\arraybackslash\hspace{0pt}}m{#1}}

\usepackage{url}
\usepackage{color}
\usepackage{blkarray} 
\usepackage{tikz}
\usepackage{pgfplots,pgfplotstable}
\usetikzlibrary{plotmarks,shapes,patterns,decorations.pathreplacing,backgrounds,calc,arrows,arrows.meta,spy,matrix,shadows,trees,positioning}
\usepgfplotslibrary{patchplots,groupplots}
\usepackage{tikzscale}
\usepackage{tikz-qtree}
\usetikzlibrary{calc}
\usepackage[export]{adjustbox} 
\usepackage{setspace}
\usepackage{optidef}

\let\oldFootnote\footnote
\newcommand\nextToken\relax

\renewcommand\footnote[1]{%
    \oldFootnote{#1}\futurelet\nextToken\isFootnote}

\newcommand\isFootnote{%
    \ifx\footnote\nextToken\textsuperscript{,}\fi}

\DeclareMathOperator*{\argmax}{\arg\!\max} 
 
\usepackage{todonotes}


{}
{}
{}
{}

\usepackage[normalem]{ulem} 

\usepackage{booktabs}

\usepackage[nopostdot,acronym,shortcuts,nonumberlist]{glossaries}
\newacronym{quic}{QUIC}{Quick UDP Internet Connections}
\newacronym{3gpp}{3GPP}{3GPP}
\newacronym{adc}{ADC}{Analog to Digital Converter}
\newacronym{5g}{5G}{5th Generation}
\newacronym{6g}{6G}{6th Generation}
\newacronym{aimd}{AIMD}{Additive Increase Multiplicative Decrease}
\newacronym{am}{AM}{Acknowledged Mode}
\newacronym{amc}{AMC}{Adaptive Modulation and Coding}
\newacronym{aqm}{AQM}{Active Queue Management}
\newacronym{awgn}{AGWN}{Additive White Gaussian Noise}
\newacronym{cvar}{CVaR}{Conditional Value at Risk}
\newacronym{balia}{BALIA}{Balanced Link Adaptation}
\newacronym{bdp}{BDP}{Bandwidth-Delay Product}
\newacronym{bbu}{BBU}{Baseband Unit}
\newacronym{bf}{BF}{Beamforming}
\newacronym{cc}{CC}{Congestion Control}
\newacronym{cdf}{CDF}{Cumulative Distribution Function}
\newacronym{ci}{CI}{Close-in free space reference}
\newacronym{cn}{CN}{Core Network}
\newacronym{cqi}{CQI}{Channel Quality Information}
\newacronym{cp}{CP}{Control Plane}
\newacronym{csirs}{CSI-RS}{Channel State Information - Reference Signal}
\newacronym{dc}{DC}{Dual Connectivity}
\newacronym{dce}{DCE}{Direct Code Execution}
\newacronym{dci}{DCI}{Downlink Control Information}
\newacronym{dl}{DL}{Downlink}
\newacronym{dmr}{DMR}{Deadline Miss Ratio}
\newacronym{dmrs}{DMRS}{DeModulation Reference Signal}
\newacronym{e2e}{E2E}{End-to-End}
\newacronym{ecn}{ECN}{Explicit Congestion Notification}
\newacronym{edf}{EDF}{Earliest Deadline First}
\newacronym{enb}{eNB}{evolved Node Base}
\newacronym{epc}{EPC}{Evolved Packet Core}
\newacronym{es}{ES}{Edge Server}
\newacronym{fdma}{FDMA}{Frequency Division Multiple Access}
\newacronym{fdd}{FDD}{Frequency Division Duplexing}
\newacronym[firstplural=Radio Access Technologies (RATs)]{rat}{RAT}{Radio Access Technology}
\newacronym{fs}{FS}{Fast Switching}
\newacronym{ftp}{FTP}{File Transfer Protocol}
\newacronym{bs}{BS}{Base Station}
\newacronym{gnb}{gNB}{Next Generation Node Base}
\newacronym{harq}{HARQ}{Hybrid Automatic Repeat reQuest}
\newacronym{hetnet}{HetNet}{Heterogeneous Network}
\newacronym{hh}{HH}{Hard Handover}
\newacronym{hol}{HOL}{Head-of-Line}
\newacronym{ia}{IA}{Initial Access}
\newacronym{rrh}{RRH}{Remote Radio Head}
\newacronym{imt}{IMT}{International Mobile Telecommunication}
\newacronym{iot}{IoT}{Internet of Things}
\newacronym{kpi}{KPI}{Key Performance Indicator}
\newacronym{los}{LOS}{Line of Sight}
\newacronym{lte}{LTE}{Long Term Evolution}
\newacronym{m2m}{M2M}{Machine to Machine}
\newacronym{mac}{MAC}{Medium Access Control}
\newacronym{mc}{MC}{Multi-Connectivity}
\newacronym{mcs}{MCS}{Modulation and Coding Scheme}
\newacronym{mec}{MEC}{Mobile Edge Cloud}
\newacronym{mi}{MI}{Mutual Information}
\newacronym{mimo}{MIMO}{Multiple Input, Multiple Output}
\newacronym{mmwave}{mmWave}{millimeter wave}
\newacronym{mr}{MR}{Maximum Rate}
\newacronym{mss}{MSS}{Maximum Segment Size}
\newacronym{mbs}{MBS}{Macro Base Station}
\newacronym{sbs}{SBS}{Small Base Station}
\newacronym{ssb}{SSB}{Synchronization Signal Block}
\newacronym{mtd}{MTD}{Machine-Type Device}
\newacronym{mtu}{MTU}{Maximum Transmission Unit}
\newacronym{nsf}{NSF}{National Science Foundation}
\newacronym{zf}{ZF}{Zero-Forcing}
\newacronym{nfv}{NFV}{Network Function Virtualization}
\newacronym{nlos}{NLOS}{Non Line of Sight}
\newacronym{nr}{NR}{New Radio}
\newacronym{ofdm}{OFDM}{Orthogonal Frequency Division Multiplexing}
\newacronym{pdcch}{PDCCH}{Physical Downlonk Control Channel}
\newacronym{pdcp}{PDCP}{Packet Data Convergence Protocol}
\newacronym{pdsch}{PDSCH}{Physical Downlink Shared Channel}
\newacronym{pdu}{PDU}{Packet Data Unit}
\newacronym{pf}{PF}{Proportional Fair}
\newacronym{pgw}{PGW}{Packet Gateway}
\newacronym{phy}{PHY}{Physical}
\newacronym{pbch}{PBCH}{Physical Broadcast Channel}
\newacronym[plural=\gls{mme}s,firstplural=Mobility Management Entities (MMEs)]{mme}{MME}{Mobility Management Entity}
\newacronym{prb}{PRB}{Physical Resource Block}
\newacronym{pss}{PSS}{Primary Synchronization Signal}
\newacronym{pucch}{PUCCH}{Physical Uplink Control Channel}
\newacronym{pusch}{PUSCH}{Physical Uplink Shared Channel}
\newacronym{rach}{RACH}{Random Access Channel}
\newacronym{ran}{RAN}{Radio Access Network}
\newacronym{red}{RED}{Random Early Detection}
\newacronym{rf}{RF}{Radio Frequency}
\newacronym{rlc}{RLC}{Radio Link Control}
\newacronym{rl}{RL}{Reinforcement Learning}
\newacronym{rlf}{RLF}{Radio Link Failure}
\newacronym{rrc}{RRC}{Radio Resource Control}
\newacronym{rrm}{RRM}{Radio Resource Management}
\newacronym{rr}{RR}{Round Robin}
\newacronym{rs}{RS}{Remote Server}
\newacronym{rsrp}{RSRP}{Reference Signal Received Power}
\newacronym{rss}{RSS}{Received Signal Strength}
\newacronym{rtt}{RTT}{Round Trip Time}
\newacronym{rw}{RW}{Receive Window}
\newacronym{rx}{RX}{Receiver}
\newacronym{sa}{SA}{standalone}
\newacronym{sack}{SACK}{Selective Acknowledgment}
\newacronym{sap}{SAP}{Service Access Point}
\newacronym{sch}{SCH}{Secondary Cell Handover}
\newacronym{scoot}{SCOOT}{Split Cycle Offset Optimization Technique}
\newacronym{sdma}{SDMA}{Spatial Division Multiple Access}
\newacronym{sinr}{SINR}{Signal to Interference plus Noise Ratio}
\newacronym{sm}{SM}{Saturation Mode}
\newacronym{snr}{SNR}{Signal to Noise Ratio}
\newacronym{son}{SON}{Self-Organizing Network}
\newacronym{ss}{SS}{Synchronization Signal}
\newacronym{srs}{SRS}{Sounding Reference Signal}
\newacronym{sss}{SSS}{Secondary Synchronization Signal}
\newacronym{tb}{TB}{Transport Block}
\newacronym{tcp}{TCP}{Transmission Control Protocol}
\newacronym{tdd}{TDD}{Time Division Duplexing}
\newacronym{tdma}{TDMA}{Time Division Multiple Access}
\newacronym{tfl}{TfL}{Transport for London}
\newacronym{thz}{THz}{Terahertz}
\newacronym{tm}{TM}{Transparent Mode}
\newacronym{trp}{TRP}{Transmitter Receiver Pair}
\newacronym{tti}{TTI}{Transmission Time Interval}
\newacronym{ttt}{TTT}{Time-to-Trigger}
\newacronym{tx}{TX}{Transmitter}
\newacronym{ue}{UE}{User Equipment}
\newacronym{ul}{UL}{Uplink}
\newacronym{uml}{UML}{Unified Modeling Language}
\newacronym{um}{UM}{Unacknowledged Mode}
\newacronym{utc}{UTC}{Urban Traffic Control}
\newacronym{vm}{VM}{Virtual Machine}
\newacronym{rsrq}{RSRQ}{Reference Signal Received Quality}
\newacronym{rssi}{RSSI}{Received Signal Strength Indicator}
\newacronym{crs}{CRS}{Cell Reference Signal}
\newacronym{comp}{CoMP}{Coordinated Multi-Point}
\newacronym{cran}{C-RAN}{Cloud \acrlong{ran}}
\newacronym{ca}{CA}{Carrier Aggregation}
\newacronym{cco}{CC}{Carrier Component}
\newacronym{nsa}{NSA}{Non Stand Alone}
\newacronym{embb}{eMBB}{Enhanced Mobility Broadband}
\newacronym{bsr}{BSR}{Buffer Status Report}
\newacronym{srb}{SRB}{Service Radio Bearer}
\newacronym{scm}{SCM}{Spatial Channel Model}
\newacronym{sctp}{SCTP}{Stream Control Transmission Protocol}
\newacronym{mptcp}{MPTCP}{Multi-path TCP}
\newacronym{ietf}{IETF}{Internet Engineering Task Force}
\newacronym{os}{OS}{Operating System}
\newacronym{tls}{TLS}{Transport Layer Security}
\newacronym{rfc}{RFC}{Request for Comments}
\newacronym{http}{HTTP}{HyperText Transfer Protocol}
\newacronym{nat}{NAT}{Network Address Translation}
\newacronym{api}{API}{Application Programming Interface}
\newacronym{rto}{RTO}{Retransmission Timeout}
\newacronym{psc}{PSC}{Public Safety Communication}
\newacronym{rpgm}{RPGM}{Reference Point Group Mobility}
\newacronym{ic}{IC}{Incident Command}
\newacronym{rsu}{RSU}{Road Side Unit}
\newacronym{uav}{UAV}{Unmanned Aerial Vehicle}
\newacronym{usa}{U.S.}{United States}
\newacronym{vr}{VR}{Virtual Reality}
\newacronym{iab}{IAB}{Integrated Access and Backhaul}
\newacronym{wlan}{WLAN}{Wireless Local Area Network}
\newacronym{cots}{COTS}{Commercial Off-the-Shelf}
\newacronym{fpga}{FPGA}{Field Programmable Gate Array}
\newacronym{rcn}{RCN}{Research Coordination Network}
\newacronym{abg}{ABG}{Alpha-Beta-Gamma}
\newacronym{fi}{FI}{Floating Intercept}
\newacronym{uas}{UAS}{Unmanned Aerial System}
\newacronym{gps}{GPS}{Global Positioning System}
\newacronym{a2g}{A2G}{air-to-ground}
\newacronym{a2a}{A2A}{air-to-air}
\newacronym{uma}{UMa}{Urban Macro}
\newacronym{umi}{UMi}{Urban Micro}
\newacronym{rma}{RMa}{Rural Macro}
\newacronym{inoo}{InOo}{Indoor Open Office}
\newacronym{ple}{PLE}{path loss exponent}
\newacronym{aoa}{AoA}{Angle of Arrival}
\newacronym{aod}{AoD}{Angle of Departure}
\newacronym{toa}{ToA}{Time of Arrival}
\newacronym{mpc}{MPC}{Multi-path Component}
\newacronym{cir}{CIR}{Channel Impulse Response}
\newacronym{rt}{RT}{Ray-tracing}
\newacronym{tc}{TC}{Time Cluster}
\newacronym{sl}{SL}{Spatial Lobe}
\newacronym{ns3}{ns-3}{Network Simulator 3}
\newacronym{fsc}{FS}{Fully Stochastic}
\newacronym{hbc}{HB}{Hybrid}
\newacronym{hpbw}{HPBW}{Half Power Beamwidth}
\newacronym{hsc}{HSC}{Hybrid Semantic Compression}
\newacronym{per}{PER}{Packet Error Rate}
\newacronym{ici}{ICI}{inter-cell interference}
\newacronym{psd}{PSD}{Power Spectrum Density}
\newacronym{du}{DU}{Distributed Unit}
\newacronym{mt}{MT}{Mobile Termination}
\newacronym{bap}{BAP}{Backhaul Adaptation Protocol}
\newacronym{mlr}{MLR}{Maximum Local Rate}
\newacronym{scaros}{SCAROS}{Scalable and Robust Self-backhauling Solution}
\newacronym{mdp}{MDP}{Markov Decision Process}

\newcommand{\name}{Safehaul}

\newfont{\mycrnotice}{ptmr8t at 7pt}
\newfont{\myconfname}{ptmri8t at 7pt}
\newcommand{\mpg}[1]{{\color{red}MP: #1}}

\newcommand{\donor}{BS$_{donor}$}
\newcommand{\donors}{BSs$_{donor}$}
\newcommand{\node}{BS$_{node}$}
\newcommand{\nodes}{BSs$_{node}$}
\IEEEoverridecommandlockouts
\allowdisplaybreaks
\begin{document}

\def\sharedaffiliation{%
\end{tabular}
\begin{tabular}{c}}


\title{\name: Risk-Averse Learning for Reliable mmWave Self-Backhauling in 6G Networks}
\author{
\IEEEauthorblockN{Amir Ashtari Gargari\IEEEauthorrefmark{1}, Andrea Ortiz\IEEEauthorrefmark{2}, Matteo Pagin\IEEEauthorrefmark{1}, Anja Klein\IEEEauthorrefmark{2}, Matthias Hollick\IEEEauthorrefmark{3}, Michele Zorzi\IEEEauthorrefmark{1}, Arash Asadi\IEEEauthorrefmark{4}}
\IEEEauthorblockA{\IEEEauthorrefmark{1}Department of Information Engineering, University of Padova, Italy, \{amirashtari, paginmatte, zorzi\}@dei.unipd.it}
\IEEEauthorblockA{\IEEEauthorrefmark{2}Communications Engineering Lab, TU Darmstadt, Germany, \{a.ortiz, a.klein\}@nt.tu-darmstadt.de}
\IEEEauthorblockA{\IEEEauthorrefmark{3}Secure Mobile Networking Lab (SEEMOO), TU Darmstadt, Germany, mhollick@seemoo.tu-darmstadt.de}
\IEEEauthorblockA{\IEEEauthorrefmark{4}Wireless Communications and Sensing Lab (WISE), TU Darmstadt, Germany, aasadi@wise.tu-darmstadt.de}}

\maketitle

\thispagestyle{empty}
\pagestyle{empty}

\IEEEtitleabstractindextext{\begin{abstract}

Wireless backhauling at millimeter-wave frequencies (mmWave) in static scenarios is a well-established practice in cellular networks. However, highly directional and adaptive beamforming in today's mmWave systems have opened new possibilities for self-backhauling. Tapping into this potential, 3GPP has standardized \gls{iab} allowing the same base station to serve both, access and backhaul traffic. Although much more cost-effective and flexible, resource allocation and path selection in \gls{iab} mmWave networks is a formidable task. To date, prior works have addressed this challenge through a plethora of classic optimization and learning methods, generally optimizing a \gls{kpi} such as throughput, latency, and fairness, and little attention has been paid to the reliability of the \gls{kpi}. We propose \name{}, a risk-averse learning-based solution for \gls{iab} mmWave networks. In addition to optimizing average performance, \name{} ensures reliability by minimizing the losses in the tail of the performance distribution. We develop a novel simulator and show via extensive simulations that \name{} not only reduces the latency by up to 43.2\% compared to the benchmarks, but also exhibits significantly more reliable performance, e.g., 71.4\% less variance in achieved latency.

\end{abstract}
}
\IEEEdisplaynontitleabstractindextext
\IEEEpeerreviewmaketitle
\glsresetall

\section{Introduction}
\label{s:introduction}

The emergence of mmWave cellular systems created a unique opportunity for cellular operators to leverage a scalable and cost-effective approach to deal with network densification. The fact that mmWave base stations can support fiber-like data rates facilitates the use of the same base station for both access and backhaul traffic, a solution which in 3GPP parlance is referred to as \gls{iab}. Consequently, 3GPP has included IAB in the standard \cite{3gpp.38.300, 3gpp_38_174} covering the details on architecture, higher layer protocols, and the radio. Although Release 17 of the 5G-NR defines the interfaces, architectures, and certain system parameters, the actual configuration and resource allocation is left to operators. 

Traditional self-backhauled networks featured fixed-wireless links decoupled from access networks with static configurations. In contrast, IAB should account for the dynamic nature of the backhaul links (particularly in dense mmWave deployments) and their integration with the access network. Further, IAB allows the traffic to traverse several hops (i.e., base stations) to reach its destination, adding a new dimension to the problem's complexity. \textit{In addition to the scheduling problem, an IAB network should: $(i)$ solve the problem of path selection and link activation at the backhaul while considering inter-cell interference and $(ii)$ decide on serving access or backhaul traffic depending on the access load and the ingress backhaul traffic from neighboring base stations.}

\textbf{Prior work.} Methodologically, the majority of the existing works~\cite{pan2017joint, alizadeh2019load, huang2015joint, nguyen2020nonsmooth, rasekh2015interference, kwon2019joint, pizzo2017optimal, kulkarni2018max,9839601} focus on classic optimization techniques to solve the above-mentioned problem. However, given the large number of parameters involved, such formulations often result in (non-)convex problems that are too complex for real-time operations. Nonetheless, they are valuable indicators to mark the upper-bound performances. Recently, some works focus on more practical solutions which can be deployed in real networks\cite{lei2020deep, ZHANG2021108248, 9473755}. Specifically, these works leverage \gls{rl} to tackle both resource allocation and/or path selection in IAB mmWave networks and demonstrate that \gls{rl}-based solutions achieve real-time performance. 

Regardless of the methodology,  prior works mostly aim at maximizing the network capacity~\cite{pan2017joint, alizadeh2019load, huang2015joint, nguyen2020nonsmooth, rasekh2015interference, kwon2019joint, pizzo2017optimal, kulkarni2018max}, minimizing latency~\cite{vu2018path, ortiz2019scaros} and improving throughput fairness~\cite{alizadeh2019load, pagin2022}. Although optimizing capacity and latency is a challenging task by itself, network operators are often more \textit{concerned about the reliability of such approaches}. This is the underlying reason that many commercial products rely on \textit{simplified} but reliable algorithms for resource allocation, despite their sub-optimal performance.". 

In this article, we propose \name{}, a reinforcement learning-based solution for reliable scheduling and path selection in IAB mmWave systems under network dynamics. We use the concept of risk aversion, commonly used in economics \cite{Rockafellar2000, Levy1998}, to measure and enhance the reliability of \name{}. The following summarizes our contributions: 

\begin{itemize}
\item We model the scheduling and path selection problem in IAB mmWave networks as a multi-agent multi-armed bandit problem (Section~\ref{s:prob_formulation}). We consider multiple fiber base stations simultaneously supporting many self-backhauled mmWave base stations. In our model, the self-backhauled base stations independently decide the links to activate. The consensus among the base stations is reached via standard-defined procedures (Section~\ref{s:consensus}).
\item We present the first solution to provide reliable performance in IAB-enabled networks (Section~\ref{s:algo}). Specifically, we investigate the joint minimization of the average end-to-end latency and its expected tail loss for each base station.  
To this aim, we propose \name{}, a learning approach that leverages a coherent risk measure called \gls{cvar}\cite{Rockafellar2000}.
\gls{cvar} measures the tail average of the end-to-end latency distribution that exceeds the maximum permitted latency, thus ensuring the reliability of the network.
\item We provide a new means of simulating multi-hop \gls{iab} networks by extending NVIDA's newly released GPU-accelerated simulator, i.e., Sionna~\cite{hoydis2022sionna} (Section~\ref{s:simulation_setup}). Specifically, we add codebook-based analog beamforming capability for both uplink and downlink communications. Further, we extend Sionna by implementing system-level components such as layer-2 schedulers and buffers and \gls{bap}-like routing across the \gls{iab} network. We believe our IAB extensions will be instrumental for the open-source evaluation of future research on self-backhauled mmWave networks.

\item Exploiting the above simulator, we evaluate and benchmark~\name{} against two of the state-of-the-art algorithms~\cite{ortiz2019scaros, polese2018distributed}. The results confirm that \name{} is significantly more reliable than benchmarks as it exhibits much tighter variance both in terms of latency (up to 71.4\%) and packet drop rate (at least 39.1\%). Further, \name{} achieves up to 43.2\% lower average latency and 11.7\% higher average throughput than the reference schemes. 
\end{itemize}

\section{System Model}
\label{s:sys_model}
We consider a cellular system with $N$ base stations capable of self-backhauling and $D$ base station with a fiber connection to the core network. Following 3GPP terminology, we refer to self-backhauled base stations as IAB-nodes (BS$_{node}$) and the fiber base stations as IAB-donors (BS$_{donor}$). 
Each IAB-node connects to the core network via a (multi-hop) wireless link to an IAB-donor. 
The sets of all BSs$_{node}$ and BSs$_{donor}$ are denoted by $\mathcal{N}=\{1,\dots,N\}$ and $ \mathcal{D}=\{1,\dots,D\}$, respectively.
The system works in a time-slotted fashion starting from time slot $i=1$ until a finite time horizon $I$. 
All the time slots $i=1,\dots,I$ have the same duration.
The IAB-nodes are equipped with two RF chains. One RF chain is used exclusively for the communication with cellular users (access network), while the second RF chain is used for self-backhauling.
In line with the 3GPP specification~\cite{3gpp_38_874}, we assume half-duplex self-backhauling, i.e., in each time slot $i$ an IAB-node can either transmit data,  receive data or remain idle. 

 We model the connections between the base stations as a graph $\mathcal{G}_i=\{\mathcal{V},\mathcal{E}_i\}$, see Fig.~\ref{fig:systModel}. The set $\mathcal{V}=\mathcal{N}\cup\mathcal{D}$ of vertices is formed by all the BS$_{node}$ and BS$_{donor}$ in the system. The set $\mathcal{E}_i$ of edges is composed of the available wireless links $(n, l)$ between a BS$_{node}$ $n\in \mathcal{N}$ and any BS (\donor{} or \node{}) $l\in \mathcal{V}$ in time slot $i$. Note that the graph $\mathcal{G}_i$ is not static. In a given time slot $i$, some links can be unavailable due to failure, blockage, or interference. Thus, only feasible wireless links are considered in the set $\mathcal{E}_i$. 
 The path $X_{n,d}$ from BS$_{node}$ $n$ to any BS$_{donor}$ $d$ is a sequence of intermediate links $(n,l)$. Note that  $X_{n,d}$ changes over time according to the traffic loads of the intermediate BS$_{node}$ and the channel conditions.
 We model the activation of link $(n,l)$ with the binary variable $x_{n,l,i}$. When $x_{n,l,i}=1$, the link is activated and BSs$_{node}$ $n$ transmits to BS $l \in \mathcal{V}$ during time slot $i$. Conversely, $x_{n,l,i}=0$ indicates the link is deactivated and no transmission can occur.

\begin{figure}[t!]
    \centering
    \includegraphics[width=0.7\columnwidth]{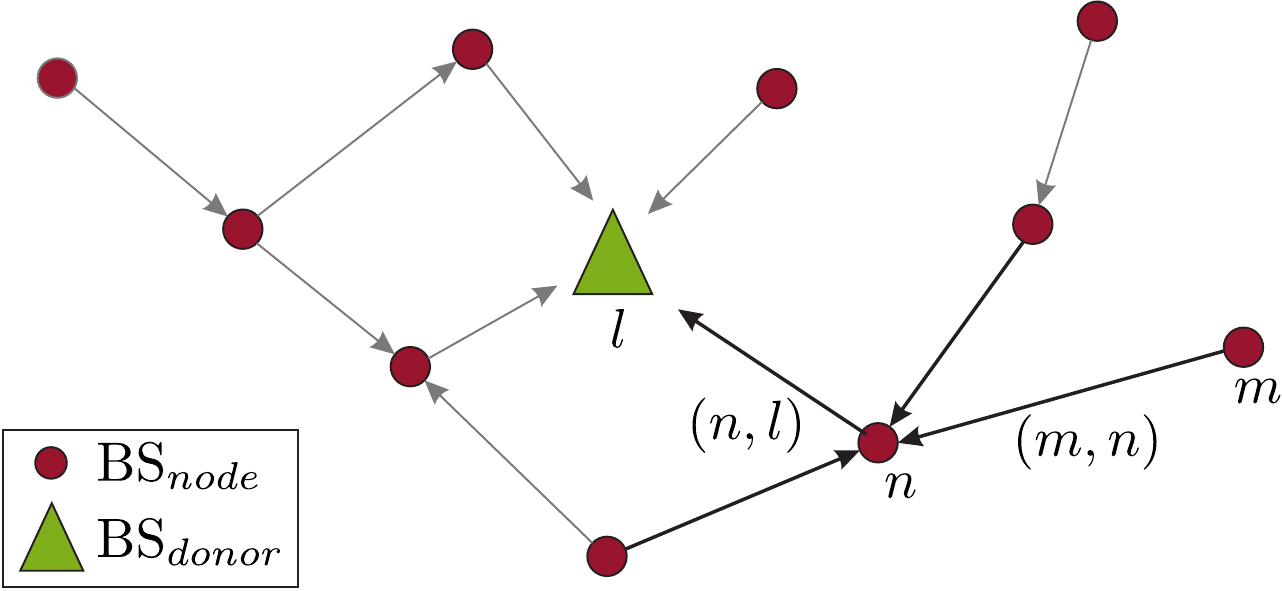}
      \caption{Example of a graph $\mathcal{G}_i$}
      \label{fig:systModel}
      \vspace{-4mm}
\end{figure}

 Each BS$_{node}$ $n$ has a finite data buffer with capacity $B^\mathrm{max}_n$ to store the backhaul data to be transmitted to any of the BSs$_{donor}$. 
 In each time slot $i$, BS$_{node}$ $n$ is characterized by its load and average queuing time. The load, denoted by $B_{n,i} \in \mathbb{N}$, indicates the number of data packets stored in the buffer at the beginning of time slot $i$. The average queuing time $t^\mathrm{q}_{n,i} \in \mathbb{R}^+$ is the average number of time slots the current packets in the data buffer have been stored.
 Additionally, we denote by $t^\mathrm{tx}_{n,l,i} \in \mathbb{R^+}$ and $M_{n,l,i} \in \mathbb{R}^+$ the transmission time from BS$_{node}$ $n$ to BS $l$ in time slot $i$, and the amount of successfully transmitted data on the link, respectively.
 We define the maximum tolerable latency $T_\mathrm{max}$ as the maximum time a packet can take from its source BS$_{node}$ to any BS$_{donor}$. Any packet that is not delivered before $T_\mathrm{max}$ milliseconds will be dropped.
 The average maximum end-to-end latency $\bar{T}_{n,d}$ from BS$_{node}$ $n$ to BS$_{donor}$ $d$ is the average, over the complete time horizon $I$, of the maximum delay a packet originating from BS$_{node}$ $n$ takes to reach any BS$_{donor}$ $d$ in time slot $i$.
 This is calculated as: 
\begin{equation}
    \bar{T}_{n,d} = \frac{1}{I}\sum_{i=1}^I T_{n,d,i}, \label{eq:avgDelayBS} 
\end{equation}
where $T_{n,d,i}$ is the maximum end-to-end latency among all the packets originating in BS$_{node}$ $n$ which reach BS$_{donor}$ $d$ in time slot $i$. 
$T_{n,d,i}$ is a sample of the random variable $T_{n,d}$ drawn from an unknown stationary probability distribution $P$ that depends on the activated links $x_{n,l,i}$, the cellular user's mobility, the location of the \node{} $n$, the interference in the system, and the queue dynamics.
  Considering \eqref{eq:avgDelayBS}, the average end-to-end latency in the system $\bar{T}$ is defined as
\begin{equation}
    \bar{T} = \frac{1}{ND}\sum_{n=1}^N\sum_{d=1}^D \bar{T}_{n,d}.
    \label{eq:avgDelay}
\end{equation}

\section{Problem Formulation}
\label{s:prob_formulation}
The joint minimization of the average end-to-end latency and the expected value of its tail loss in IAB-enabled networks is formulated in this section. We first introduce \gls{cvar}, the risk metric accounting for minimizing the events in which the end-to-end latency is higher than $T_\mathrm{max}$. Next, we formulate the optimization problem in the complete network.
 
\subsection{Preliminaries on \gls{cvar}}
Traditionally, latency minimization in IAB-enabled networks has focused on optimizing the expected value of a latency function~\cite{vu2018path, ortiz2019scaros}.
However, such an approach fails to capture the time variability of the latency distribution, thus leading to unreliable systems in which $T_{n,d,i}>T_\mathrm{max}$, for any  $i=1,...,I$, $n\in \mathcal{N}$ and $d\in\mathcal{D}$.
\textit{For this purpose, we consider not only the average end-to-end latency $\bar{T}$ in the system, but also its expected tail loss based on the \gls{cvar}}\cite{Rockafellar2000, Rockafellar2002}.

Having in mind that $T_{n,d}$ is a random variable, we assume it has a bounded mean on a probability space $(\Omega, \mathcal{F}, P)$, with $\Omega$ and  $\mathcal{F}$ being the sample and event space, respectively.
Using a risk level $\alpha\in(0,1]$, the $\mathrm{\gls{cvar}}_\alpha(T_{n,d})$ of $T_{n,d}$ at risk level $\alpha$ quantifies the losses that might be encountered in the $\alpha$-tail. More specifically, it is the expected value of $T_{n,d}$ in its $\alpha$-tail distribution \cite{Rockafellar2002}. 
Formally, $\mathrm{CVaR}_\alpha(T_{n,d})$ is defined as \cite{Rockafellar2000}
\begin{equation}
    \mathrm{CVaR}_\alpha(T_{n,d}) = \min_{q\in\mathbb{R}}\left\{q+\frac{1}{\alpha}\mathbb{E}\left[\max\{T_{n,d}-q,0\}\right]\right\},
    \label{eq:CVaR}
\end{equation}
where the expectation in \eqref{eq:CVaR} is taken over the probability distribution $P$. 
Note that lower $\mathrm{\gls{cvar}}_\alpha(T_{n,d})$ results in higher reliability in the system because the expected end-to-end latency in the $\alpha$-worst cases is low.
Moreover, note that $\alpha$ is a risk aversion parameter. For $\alpha=1$,  $\mathrm{\gls{cvar}}_\alpha(T_{n,d}) = \mathbb{E}[T_{n,d}]$ which corresponds to the traditional risk-neutral case. 
Conversely, for $\alpha=0$,  $\lim_{\alpha\rightarrow 0}\mathrm{\gls{cvar}}_\alpha(T_{n,d}) = \sup\{T_{n,d}\}$.
\gls{cvar} has been shown to be a coherent risk measure, i.e., it fulfills monotonicity, subadditivity, translation invariance, and positive homogeneity properties \cite{Pflug2000}. 

\subsection{Optimization Problem}
We aim to jointly minimize the average end-to-end latency and its expected tail loss for each \node{}. 
For this purpose, we decide which of the $(n,l)$ links to activate in each time slot $i$ during the finite time horizon $I$.
In the following, we formulate the optimization problem from the network perspective and consider the sum over all the \node{} in the system. 
The latency minimization problem should consider three different aspects: $(i)$ link activation is constrained by the half-duplex nature of self-backhauling, $(ii)$ only data stored in the data buffers can be transmitted, and $(iii)$ packet drop due to buffer overflow should be avoided.
Formally, the problem is written as:
\begin{mini!}[3]
		{\{x_{n,l,i}, i\in[1,I]\}}{\sum_{n\in\mathcal{N}}\sum_{d\in\mathcal{D}}\bar{T}_{n,d}+\eta\mathrm{\gls{cvar}}_\alpha(T_{n,f}) \label{eq:objective}}{\label{eq:OptProblem}}{}
		\addConstraint{\sum_{l\in \mathcal{V}}x_{n,l,i}+\sum_{l\in \mathcal{N}}x_{l,n,i}}{=1,\, \label{eq:constHD}}{n\in\mathcal{N}}
		\addConstraint{\sum_{j=1}^i B_{n,i}}{\geq \sum_{j=1}^i M_{n,l,i,\, \label{eq:constCausal}}}{ n\in\mathcal{N}, l\in\mathcal{V}}
		\addConstraint{\sum_{j=1}^i B_{l,j} + \sum_{j=1}^i M_{n,l,j}}{\leq B^\mathrm{max}_l,\,}{ n\in\mathcal{N}, l\in\mathcal{V}}
		\addConstraint{x_{n,l,i}}{\in \{0,1\}, \label{eq:constOverflow}}{n\in \mathcal{N}, l\in \mathcal{V}.}
\end{mini!}
In \eqref{eq:objective}, $\eta \in [0,1]$ is a weighting parameter to trade between minimizing the average end-to-end latency $\bar{T}_{n,d}$ and the expected loss of its tail.
As the considered scenario is not static, solving \eqref{eq:OptProblem} requires complete non-causal knowledge of the system dynamics during the complete time horizon $I$. However, in practical scenarios, knowledge about the underlying random processes is not available in advance.
For example, the IAB-node's loads $B_{n,i}$ depend not only on the transmitted and received backhaul data, but also on the randomly arriving data from its users. Similarly, the amounts of transmitted data $M_{n,l,i}$ depend on the varying channel conditions of both BS $n$ and $l$. 
As a result, the exact values of $T_{n,l,i}$, $B_{n,i}$ and $M_{n,l,i}$ are not known beforehand.
For this reason, we present in Sec. \ref{s:algo} \name{}, a multi-agent learning approach to minimize in each \node{} the average end-to-end latency and the expected value of the tail of its loss.
 

\section{Our proposed solution: \name}
\label{s:algo}
In this section, we describe \name{}, a multi-agent learning approach for the joint minimization of the average end-to-end latency and its expected tail loss in IAB mmWave networks. 
In \name{}, each \node{} independently decides which links $(n,l)$ to activate in every time slot $i$ by leveraging a multi-armed bandit formulation. The consensus among the \node{} is reached by exploiting the centralized resource coordination and topology management role of \gls{iab}-donors~\cite[Sec. 4.7.1]{3gpp.38.300}. 

\subsection{Multi-Armed Bandit Formulation}
Multi-armed bandits is a tool well suited to problems in which an agent makes sequential decisions in an unknown environment\cite{Sutton2018}. 
In our scenario, each \node{} $n$ decides, in each time slot $i$, which of the links $(n,l)$ to activate without requiring prior knowledge about the system dynamics.
The multi-armed bandit problem at \node{} $n$ can be characterized by a set $\mathcal{A}_n$ of actions and a set $\mathcal{R}_n$ of possible rewards. 
The rewards $r_{n,i} \in \mathcal{R}_n$ are obtained in each time slot $i$ as a response to the selected action $a_{n,i} \in \mathcal{A}_n$.
Specifically, the actions are the links that \node{} $n$ can activate, and the rewards are a function of the observed latency.
We define $\mathcal{A}_n$ as
\begin{equation}
\mathcal{A}_{n} = \{(n,l), (m,n)| n,m\in \mathcal{N}, \,l\in\mathcal{V} \},  
\end{equation}
where link $(n,n)$ indicates the \node{} $n$ remains idle.
As blockages, overloads, or failures might render certain links $(n,l)$ temporarily unavailable, we define the set $\mathcal{A}_{n,i} \subseteq \mathcal{A}_n$ of available actions  in time slot $i$ as 
\begin{equation}
\mathcal{A}_{n,i} = \{(n,l), (l,n)|(n,l), (l,n) \in \mathcal{E}_i \}.  
\end{equation}
Selecting action $a_i=(n,l)$ in time slot $i$ implies $x_{n,l,i}=1$.

The rewards $r_{n,i}$ are a function of the end-to-end latencies ${T}_{n,d,i}$ and depend on whether at \node{} $n$ a link $(n,l)$ or $(l,n)$ is activated. 
\node{} $n$ is connected to the \donor{} via multi-hop wireless links. 
Consequently, $T_{n,d,i}$ cannot be immediately observed when a link $(n,l)$,  with $l \notin \mathcal{D}$ is activated.
In fact, the destination \donor{} $d$ might not even be known to \node{} $n$ at time slot $i$.
To overcome this limitation, we define the rewards $r_{n,i}$ as a function of the next-hop's estimated end-to-end latency $\hat{T}_{l,d,i}$ as
\begin{equation}
r_{n,i}=\begin{cases}
t^\mathrm{q}_{l,i} + t^\mathrm{tx}_{n,l,i} + \hat{T}_{l,d,i}, &\text{for link } (n,l) \\
t^\mathrm{q}_{n,i} + \hat{T}_{n,d,i}, & \text{for link } (l,n),
\end{cases}
\label{eq:rewards}
\end{equation}
where $\hat{T}_{l,d,i}$ is calculated as
\begin{equation}
   \hat{T}_{l,d,i} = \underset{(l,m) \in \mathcal{E}_i}{\min}\;\hat{T}_{l,m,i}.
   \label{eq:estimatedDelay}
\end{equation}

\subsection{Latency and \gls{cvar} Estimation}
As given in \eqref{eq:rewards} and \eqref{eq:estimatedDelay}, \node{} $n$ learns which links $(n,l)$ to activate by building estimates of the expected latency $\hat{T}_{n,l}$ associated to each of them.
Let $K_{n,l,i}=\sum_{j=1}^i x_{n,l,i}$ be the number of times link $(n,l)$ has been activated up to time slot $i$. $\hat{T}_{n,l}$ is updated using the sample mean as
\begin{equation}
    \hat{T}_{n,l,i+1} = \frac{K_{n,l,i}\hat{T}_{n,l,i} + r_{n,i}}{K_{n,l,i}+1},
    \label{eq:delayEstimateSM}
\end{equation}
where the subindex $i$ is introduced to emphasize that the estimate is built over time.
The \gls{cvar} definition given in \eqref{eq:CVaR} requires $T_{n,d}$ which, as discussed before, is known a priori. 
Hence, we leverage the non-parametric estimator derived in \cite{Chen2007} to estimate the \gls{cvar} of a link $(n,l)$. To this aim, let $\Tilde{r}_n^1,\dots,\Tilde{r}_n^{K_{n,l,i}}$ be all the rewards received up to time $i$ ordered in a descending fashion, i.e., $\Tilde{r}_n^1\geq\dots\geq \Tilde{r}_n^{K_{n,l,i}}$.
The estimated $\widehat{\mathrm{\gls{cvar}}}_i(n,l)$ at time slot $i$ is calculated as \cite{Chen2007}
 \begin{equation}
     \widehat{\mathrm{\gls{cvar}}}_i(n,l)=\frac{1}{\lceil \alpha K_{n, l,i} \rceil}\sum_{k=1}^{\lceil \alpha K_{n,l,i}\rceil}{\Tilde{r}_n^k}.
     \label{eq:cvarEstimate}
 \end{equation}

Using the estimates in \eqref{eq:delayEstimateSM} and \eqref{eq:cvarEstimate}, \node{} computes the value $Q_n(a_{n,i}=(n,l))$ associated to the selected action $a_n\in \mathcal{A}_n$, and defined as
\begin{equation}
    Q_n(a_{n,i})=\hat{T}_{n.l,i} + \eta \widehat{\mathrm{\gls{cvar}}}_i(n,l).
    \label{eq:Qvalue}
\end{equation}
Note that \eqref{eq:Qvalue} is aligned with the objective function in \eqref{eq:objective}. Actions with an associated low value $Q_n(a_{n,i})$ lead to lower end-to-end latency and a low expected value on its tail.


\subsection{Consensus}
\label{s:consensus}
All the \node{} independently decide which links to activate based on their estimates of the end-to-end latency. 
As a consequence, conflicting actions may be encountered. 
A conflict occurs when two or more \node{} $n$ and $m$ aim at activating a link to a common BS $l$, $l \in \mathcal{V}$, i.e., $x_{n,l,i}=x_{m,l,i}=1$.
We reach consensus by first retrieving the buffer and congestion status of the various \gls{iab}-nodes, leveraging the related \gls{bap} layer functionality~\cite[Sec. 4.7.3]{3gpp.38.300}.
With this information at hand, conflicts are resolved by prioritizing the transmission of the \node{} with the larger queuing times $t^\mathrm{q}_{n,i}$ and loads $B_{n,i}$.
Then, we let the \gls{iab}-donor mark as \textit{unavailable} the time resources of the remaining base stations with conflicting scheduling decisions~\cite[Sec. 10.9]{3gpp.38.300}.
Note that as the learning is performed at each \node{}, only the link activation decision and the weighted sum of $t^\mathrm{q}_{n,i}$ and $B_{n,i}$ are transmitted.
Thus, low communication overhead is maintained.

\subsection{Implementation of \name{}}
Here, we describe how the above-mentioned solution can be implemented in a real system. Specifically, we elaborate on the required inputs and the interactions among the different entities as well as the pseudo-code of \name{}, see Alg. \ref{alg:algo}. 

\name{} is executed at each \node{} $n$. For its implementation, the network operator  provides $\alpha$, $\eta$ and $\mathcal{A}_n$ as an input. $\alpha$ is the risk level parameter that influences the level of reliability achieved in the system. Similarly, $\eta$ controls the impact the minimization of the latency in the $\alpha$-worst cases has on the overall performance. Both parameters, $\alpha$ and $\eta$, are set by the network operator depending on its own reliability requirements.
The set $\mathcal{A}_n$ depends on the considered network topology which is perfectly known by the network operator. The set $\mathcal{A}_n$ includes all links $(n,l)$ and $(l,n)$ to and from the first-hop neighbors of \node{} $n$.
\begin{algorithm}[t]
\footnotesize
\begin{algorithmic}[1]
    \Require $\alpha$, $\eta$, $\mathcal{A}_n$
    \State Initialize $\hat{T}_{n,l}$, $\widehat{\mathrm{\gls{cvar}}}(n,l)$, and $Q_n$ for all $(n,l) \in \mathcal{E}_1$ \label{line:initEst}
    \State Set counters $K_{n,l}=0$ and initial action $a_{n,1}=(n,n)$ \label{line:initCounter}
	\For {every time slot $i=1,...,I$}
	    \State perform action $a_{n,i}$ and observe reward $r_{n,i}$ \Comment{Eq. \eqref{eq:rewards}} \label{line:reward}
	    \State increase counter $K_{n,l}$ by one \label{line:incCounter}
		\State update latency estimate $\hat{T}_{n,l}$ \Comment{Eq: \eqref{eq:delayEstimateSM}} \label{line:updateEstDelay}
		\State update \gls{cvar} estimate $\widehat{\mathrm{\gls{cvar}}}(n,l)$ \Comment{Eq: \eqref{eq:cvarEstimate}} \label{line:updateCvar}
		\State update $Q_n(a_{n,i})$ \Comment{Eq: \eqref{eq:Qvalue}} \label{line:updateQ}
		\State select next action $a_{n,i+1}$ using $\epsilon$-greedy \Comment{Eq. \eqref{eq:epsGreedy}} \label{line:epsGreedy}
		\State share $a_{n,i+1}$, $t^\mathrm{q}_{n,i}$ and $B_{n,i}$ with the others \node{} \label{line:share}
		\State if required, update $a_{n,i+1}$ to reach consensus \Comment{Sec. \ref{s:consensus}} \label{line:consensus}
	\EndFor
\end{algorithmic}
\caption{\name{} algorithm at each \node{}} \label{alg:algo} 
\end{algorithm}

The execution of \name{} begins with the initialization of the latency and \gls{cvar} estimates, and the values $Q$ of the actions in $\mathcal{A}_n$.
Additionally, the counters $K_{n,l}$, that support the calculations of $\hat{T}_{n,l}$ and $\widehat{\mathrm{\gls{cvar}}}(n,l)$, are initialized for all links in $\mathcal{A}_n$ (line \ref{line:initEst}-\ref{line:initCounter}).
These parameters are updated and learnt throughout the execution of \name{}.
At time slot $t=0$, no transmission has occurred and $B_{n,0}=0$. Hence, \node{} $n$ remains idle for the first time slot $i=1$, i.e., $a_{n,1}=(n,n)$ (line \ref{line:initCounter}).
Next, and in all the subsequent time slots $i\in [1,I]$, the selected action is performed and the corresponding reward is obtained (line \ref{line:reward}).
If \node{} $n$ transmits in time slot $i$, i.e., $a_{n,i}=(n,l)$, the reward $r_{n,i}$ is sent by the receiving BS $l$ through the control channel.
If $a_{n,i}=(l,n)$, the reward $r_{n,i}$ depends, as given in \eqref{eq:rewards}, only on the current estimates at \node{} $n$ and the status of its buffer $B_{n,i}$.
With the observed reward $r_{n,i}$, the counter for action $a_{n,i}$ is increased and the latency and \gls{cvar} estimates are updated (lines \ref{line:incCounter}-\ref{line:updateCvar}).
Using the new estimates (lines \ref{line:updateEstDelay} and $\ref{line:updateCvar}$), the value $Q(a_{n,i})$ of the performed action $a_{n,i}$ is updated (line \ref{line:updateQ}).
The next action $a_{n,i+1}$ is then selected according to $\epsilon$-greedy (line \ref{line:epsGreedy}).
$\epsilon$-greedy is a well-known method to balance the exploitation of links with estimated low latency, and the exploration of unknown but potentially better ones.
In $\epsilon$-greedy a random action $a_{n,i+1}$ from the set $\mathcal{A}_{n,i}$ is selected with probability $\epsilon \in [0,1]$.
With probability $(1-\epsilon)$, the action that yields the estimated lowest value is chosen, i.e., 
\begin{equation}
a_{n,i+1} = \begin{cases}
\text{randomly selected action from } \mathcal{A}_{n,i}, & \text{if } x\leq\epsilon \\
\underset{b_n \in \mathcal{A}_{n,i}}{\argmax} \, Q_n(b_{n}), & \text{if } x>\epsilon, 
\end{cases}
    \label{eq:epsGreedy}
\end{equation}
where $x$ is a sample taken from a uniform distribution in the interval $[0,1]$.
Once the action $a_{n,i+1}$ is selected, it is shared with other \node{} in the network along with $t^\mathrm{q}_{n,i}$ and $B_{n,i}$ (line \ref{line:share}).
As described in Section~\ref{s:consensus}, this goes through the control channel. 
If conflicts arise, consensus is reached by prioritizing the transmission of \node{} with the largest loads and queuing times (line \ref{line:consensus}).

The regret of \name{} is defined as the expected loss caused by the fact that the optimal action is not always selected \cite{Auer2002}. For brevity, we omit the regret analysis of \name{}. Nevertheless, a regret bound can be derived following an approach similar to the work in~\cite{pmlr-v29-Galichet13} but including the $\epsilon$-greedy considerations~\cite{Auer2002}. Moreover,  the bound should account for the fact that the probability of not selecting an optimal action also depends on the actions of the other \nodes{}.


\section{Simulation setup}
\label{s:simulation_setup}
Given the lack of access to actual 5G (and beyond) network deployments, prior works mostly rely on \textit{home-grown} simulators for performance evaluation. Although a valid approach, these simulators often cannot fully capture the real network dynamics, introducing strong assumptions in the physical and/or the upper layers of the protocol stack. 
Until very recently, the most complete  simulator for \gls{iab} networks was a system-level simulator~\cite{8514996} developed as an extension of the ns-3 \textit{mmWave} module~\cite{8344116}. Despite accurate modeling of the \gls{iab} protocol stack, it is currently behind the latest \gls{iab} specifications\footnote{For instance due to the assumption of layer-3 (instead of layer-2) relaying at the \gls{iab}-nodes which was based on a draft version of the TR 38.874~\cite{3gpp_38_874_old}.}. Moreover, the ns-3 \gls{iab} extension is unsuitable for large simulations with hundreds of nodes due to reliance on an older version of the \textit{mmWave} module. Therefore, in our work we opt for Sionna~\cite{hoydis2022sionna}, which is an open-source GPU-accelerated toolkit based on TensorFlow. The tensor-based implementation supports the native integration of neural networks and prototyping complex communication systems. 

However, unlike the aforementioned ns-3 module, Sionna is a physical layer-focused simulator that does not explicitly model 5G networks, thus lacking  the characterization of the 5G-NR upper-layer protocol stack. Hence, we extend Sionna by including the system-level functionalities such as MAC-level scheduling and RLC-level buffering. Furthermore, since Sionna exhibits slight differences compared to the 5G-NR physical layer, we extend Sionna's physical layer model~\cite{hoydis2022sionna} with the 5G-NR procedures.  All these contributions will be made publicly available upon publication of this article. In the following, we describe the details of our extensions.

\begin{figure}[t!]
    \centering
    \includegraphics[scale=0.2]{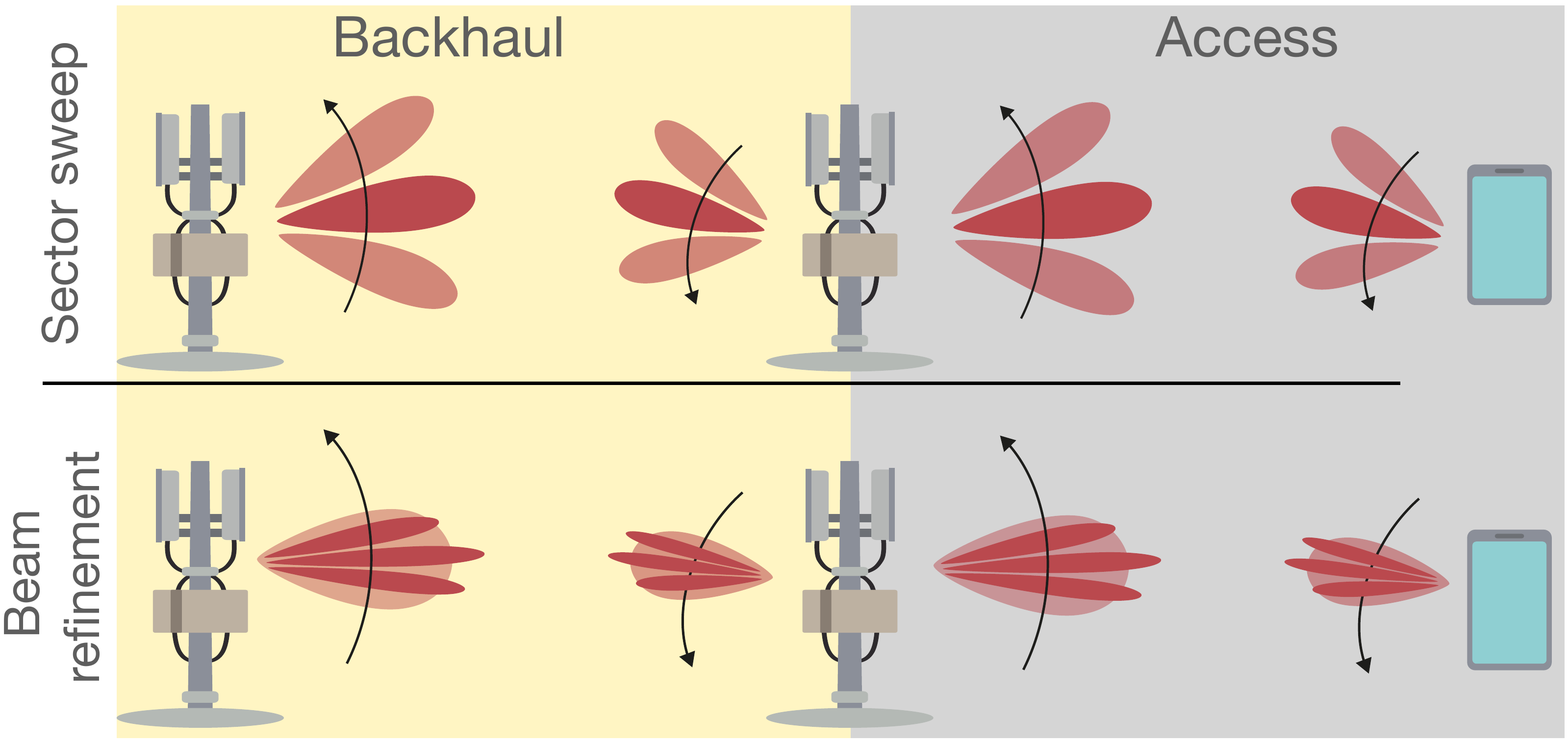}
      \caption{Schematic of the hierarchical beam management procedure. First, the general direction is estimated using wide beams (top). Then, the search is refined using the narrow beams codebook.}
      \label{fig:beamforming}
      \vspace{-5mm}
\end{figure}

\subsection{Extensions to Sionna's physical layer module}
\label{sub:additionalSionna}
In this section, we describe the physical layer modification that were necessary to evaluate IAB scenarios using Sionna. 


\subsubsection{Codebook-based Beamforming}
Sionna's native beamforming only supports \gls{zf} pre-coding in downlink. Therefore, as a first step, we extend Sionna by implementing an NR-like codebook-based analog beamforming both at the transmitter and at the receiver.
Specifically, we assume that the beamforming vectors at the transmitter $w_{tx}$ and at the receiver $w_{rx}$ are a pair of codewords selected from a predefined codebook. The codebook is computed by defining a set of beam directions $\{ \omega_{n, m} \}$ which scans a given angular sector with a fixed beamwidth. The steering vector $a_{n,m}$ corresponding to direction $\omega_{n, m}$can be computed~as:
\medmuskip=1mu
\thinmuskip=1mu
\thickmuskip=1mu
\begin{equation}
\begin{aligned}
    a_{n, m} & =  \left[ 1,\ldots, e^{j\frac{2\pi}{\lambda}d\left(i_{ H}\sin\alpha_n\sin\beta_m+i_{V}\cos\beta_m\right)}, \ldots, \right. \\ 
    & \left. e^{j\frac{2\pi}{\lambda}d\left((N_{H}-1)\sin\alpha_n\sin\beta_m+(N_{ V}-1)\cos\beta_m\right)}\right]^T,
\end{aligned}
\end{equation}
\medmuskip=6mu
\thinmuskip=6mu
\thickmuskip=6mu
where $N_{H}$ and $N_{V}$ are the number of horizontal and vertical antenna elements, respectively. The horizontal and vertical index of a radiating element is denoted by $i_{H}\in [0, N_{H}]$ and $i_{V}\in [0, N_{V}]$, respectively. $\alpha_n$ and $\beta_m$ represent the azimuth and elevation angles of $\omega_{n, m}$. Next, we define the codebook as the set $\{ \left( \sqrt{N_{H} N_{V}} \right)^{-1} a_{n, m} \}$. 

In line with the 5G-NR beam management procedure~\cite{giordani2018tutorial}, we assume the lack of complete channel knowledge, i.e., the communication endpoints do not know the corresponding channel matrix. Accordingly, an exhaustive search is conducted to identify the best pair of codewords resulting in the highest \gls{sinr}. Specifically, we leverage a hierarchical search~\cite{hosoya2014multiple}, in which the communication pairs first perform a wide-beam search (a.k.a. sector-level sweep) in which the transmitter and receiver approximate the direction of communication, see Fig.~\ref{fig:beamforming}. Next, the beamforming direction is fine-tuned through a beam refinement procedure going through a codebook with narrow beams. Consequently, we employ two types of codebooks, one with wide beams for sector sweep and another with narrow beams for beam refinement.   

\subsubsection{SINR Computations}

Since Sionna does not natively calculate the \gls{sinr}, we add this functionality to the simulator to better model the impact of interference in our simulations. We compute \gls{sinr} experienced by \glspl{tb} by combining the power of the intended signal with that of the interferers and the thermal noise. Specifically, we first compute the power $P_{i}$ of the intended signal at receiver $i$ over frequency $f$ and at time slot $t$ . Then, we obtain the overall interference power by leveraging the superposition principle and summing the received power from all other interfering base stations $P_{k} (t, f)$ where $k\in \mathbb{N}$ and $ k \neq i$ . For the purposes of this computation, we assume that each interferer employs the beamforming vector yielding the highest \gls{sinr} towards its intended destination. Similarly, the transmitter and receiver use the beamforming configuration estimated via the hierarchical search procedure. Finally, the \gls{sinr} is $
\label{EQ_CGAN1}
\gamma_{i} (t, f)= \frac{P_{i}(t,f)}{\sum\limits_{i\in \mathbb{N},k \neq i} P_{k}(t,f) + \sigma(t,f)}
$
where $\sigma(t,f)$ is the thermal noise at the receiver.

%
%
%


\usetikzlibrary{positioning,shapes,shadows,arrows}
\definecolor{mycolor}{rgb}{1,0.2,0.2}
\definecolor{mycolor2}{rgb}{0.3,0.7,0.1}
\definecolor{mycolor3}{rgb}{0.9,0.9,0.9}
\tikzstyle{abstract}=[rectangle, draw=black, rounded corners, fill=mycolor, drop shadow,
        text centered, anchor=north, text=white, text width=2.5cm]
\tikzstyle{comment}=[rectangle, draw=black, rounded corners, fill=mycolor3, drop shadow,
        text centered, anchor=north, text=black, text width=2.5cm]
\tikzstyle{Sio}=[rectangle, draw=black, rounded corners, fill=mycolor2, drop shadow,
        text centered, anchor=north, text=white, text width=2.5cm]
\tikzstyle{myarrow}=[->, >=open triangle 90, thick]
\tikzstyle{line}=[-, thick]
        
\begin{figure}\label{bigpicture}\centering
\resizebox{.45\textwidth}{!}{
\begin{tikzpicture}[node distance=2cm,thick,scale=1, every node/.style={scale=1}]
    \node (Item) [abstract, rectangle split parts=2]
        {
            \textbf{Simulator}
        };
    \node (AuxNode01) [text width=4cm, below=of Item, xshift=1cm] {};
    \node (Component) [abstract, rectangle split, rectangle split parts=2,  below=of Item, xshift=1.5cm, yshift=0.7cm]
        {
        
            \textbf{nonPHY}
            \nodepart{second} Sec. \ref{sub:nonPHY_simulator}
        };
    \node (class) [abstract, rectangle split, rectangle split parts=2, left=of Component, xshift=-0.5cm]
        {
            \textbf{DataTypes}
            \nodepart{second} sub Class()
        };
    \node (System) [abstract, rectangle split, rectangle split parts=2, right=of Component ,xshift=2cm]
        {
            \textbf{PHY}
            \nodepart{second} Physical
        };
    \node (AuxNode02) [text width=0.5cm, below=of Component] {};
    \node (AuxNode10) [text width=0.5cm, below=of class] {};
     \node (iab) [abstract, rectangle split, rectangle split parts=2, below=of class, xshift=0cm, yshift=1.4cm]
        {
            \textbf{iab()}
            \nodepart{second}DataType
        }; 

    \node (iabInstants) [comment, rectangle split, rectangle split parts=2, below=0.2cm of iab, text justified]
        {
            \textbf{Instants}
            \nodepart{second}iab\_id \newline cell\_id \newline location \newline Antenna \& BF  \newline Tx\_Buffer \newline Rx\_Buffer 
        }; 
     \node (network) [abstract, rectangle split, rectangle split parts=2, left=of class, xshift=1.5cm]
        {
            \textbf{Setup()}
            \nodepart{second}Class
        }; 

    \node (netInstants) [comment, rectangle split, rectangle split parts=2, below=0.2cm of network, text justified]
        {
            \textbf{Instants}
            \nodepart{second} Topology() \newline simulation\_mode \newline run\_time \newline Access() \newline Backhaul()
        }; 
    \node (packet) [abstract, rectangle split, rectangle split parts=2, below=of netInstants, yshift=1.3cm]
        {
            \textbf{packet()}
            \nodepart{second}DataType
        }; 

    \node (packetInstants) [comment, rectangle split, rectangle split parts=2, below=0.2cm of packet, text justified]
        {
            \textbf{Instants}
            \nodepart{second}packet\_id \newline packet\_size \newline frame\_id \newline UE\_id \newline Cell\_id \newline Delay \newline passed\_path\_list 
        };

    \node (path) [abstract, rectangle split, rectangle split parts=2, below=of iabInstants, yshift=1.7cm]
        {
            \textbf{path()}
            \nodepart{second}DataType
        }; 

    \node (pathInstants) [comment, rectangle split, rectangle split parts=2, below=0.2cm of path, text justified]
        {
            \textbf{Instants}
            \nodepart{second}path\_id \newline from\_IAB \newline to\_IAB \newline SINR \newline path\_rate \newline path\_equipped 
        };

    \node (Sensor) [abstract, rectangle split, rectangle split parts=2, left=of AuxNode02, xshift=1.8cm]
        {
            \textbf{DataFlow}
            \nodepart{second} Sec.~\ref{sub:Dataflow}
        };

    \node (Part) [abstract, rectangle split, rectangle split parts=2, right=of AuxNode02, xshift=-1.8cm]
        {
            \textbf{Control Plane}
            \nodepart{second}
        };
        

    \node (Temperature) [abstract, rectangle split, rectangle split parts=2, below=of Sensor, xshift=0cm, yshift=1cm]
        {
            \textbf{Buffer}
            \nodepart{second}RLC-like
        };
    \node (TemperatureInstants) [comment, rectangle split, rectangle split parts=2, below=0.2cm of Temperature, text justified]
        {
            \textbf{Instants}
            \nodepart{second}RX DU RLC \newline TX MT RLC
        };

    \node (Pump) [abstract, rectangle split, rectangle split parts=2, below=of Part, xshift=0cm,yshift=1cm]
        {
            \textbf{BAP-like}
            \nodepart{second} Sec.~\ref{sub:bap}
        };
    \node (PumpInstants) [comment, rectangle split, rectangle split parts=2, below=0.2cm of Pump, text justified]
        {
            \textbf{Instants}
            \nodepart{second}3GPP Standard TS 38.340
        };
     \node (pathsel) [abstract, rectangle split, rectangle split parts=2, below=of PumpInstants, xshift=0cm,yshift=1.7cm]
        {
            \textbf{Path Selection Policy}
            \nodepart{second}
        };    

    \node (pathInstants) [comment, rectangle split, rectangle split parts=2, below=0.2cm of pathsel, text justified]
        {
            \textbf{Instants}
            \nodepart{second} \name{} \newline SCAROS \newline MLR
        };    
    \node (Valve) [abstract, rectangle split, rectangle split parts=2, right=of pathInstants, xshift=0cm, yshift=1.85cm]
        {
            \textbf{Scheduler}
            \nodepart{second}Sec.~\ref{sub:scheduler}
        };

    \node (ValveInstants) [comment, rectangle split, rectangle split parts=2, below=0.2cm of Valve, text justified]
        {
            \textbf{Instants}
            \nodepart{second} RoundRobin()
        };

    \node (AuxNode05) [below=of System] {};
    \node (CoolingSystem) [Sio, rectangle split, rectangle split parts=2, left=of AuxNode05, xshift=2cm]
        {
            \textbf{Sionna}
            \nodepart{second} Sec.~\ref{s:simulation_setup}
        };
    \node (CoolingLoop) [abstract, rectangle split, rectangle split parts=2, right=of AuxNode05, xshift=-2cm]
        {
            \textbf{Extended Modules}
            \nodepart{second} Sec.~\ref{sub:additionalSionna}
        };
    \node (CoolingSystemInstants) [comment, rectangle split, rectangle split parts=2, below=0.2cm of CoolingSystem, text justified]
        {
            \textbf{Instants}
            \nodepart{second}
            \newline 3GPP TR 38.901 Channel models
            \newline MU-MIMO
            \newline OFDM
            \newline 5G-NR

        };
    \node (CoolingLoopInstants) [comment, rectangle split, rectangle split parts=2, below=0.2cm of CoolingLoop, text justified]
        {
            \textbf{Instants}
            \nodepart{second}Beamforming\newline ICI \& SINR
        };
    
    \draw[myarrow] (Component.north) -- ++(0,0.8) -| (Item.south);
    \draw[line] (Component.north) -- ++(0,0.8) -| (System.north);
    \draw[line] (network.north) -- ++(0,0.8) -| (Item.south);
    \draw[myarrow] (class.north) -- ++(0,0.8) -| (Item.south);

    \draw[myarrow] (Sensor.north) -- ++(0,0.8) -| (Component.south);
    \draw[line] (Sensor.north) -- ++(0,0.8) -| (Part.north);

    \draw[line] (Pump.east) -- ++(0.2,0);
     ([xshift=-1cm]Sensor.south);
     \draw[myarrow] (Temperature.north) -- ++(0,0.8) -| (Sensor.south);

    \draw[line] (iab.west) -- ++(-0.2,0);
    \draw[line] (packet.east) -- ++(0.3,0);
    \draw[myarrow] (path.west) -- ++(-0.2,0) -- ([yshift=-0.5cm, xshift=-0.2cm] class.south west) -|([xshift=0cm]class.south);
    

     ([xshift=-1cm]Part.south);
    \draw[myarrow] (pathsel.east) -- ++(0.2,0) -- ([yshift=0.5cm, xshift=0.2cm] Pump.north east) -|
     ([xshift=0cm]Part.south);
     
     \draw[line] (pathsel.east) -- ++(0,0.) -| (Valve.west);
    \draw[myarrow] (CoolingSystem.north) -- ++(0,0.8) -| (System.south);
    \draw[line] (CoolingSystem.north) -- ++(0,0.8) -| (CoolingLoop.north);

\end{tikzpicture}
}
\caption{Overall design of our Sionna's extension. The red blocks represent our additions to the baseline simulator, i.e., Sionna~\cite{hoydis2022sionna}.}
\label{fig:simulator_design}
\end{figure}
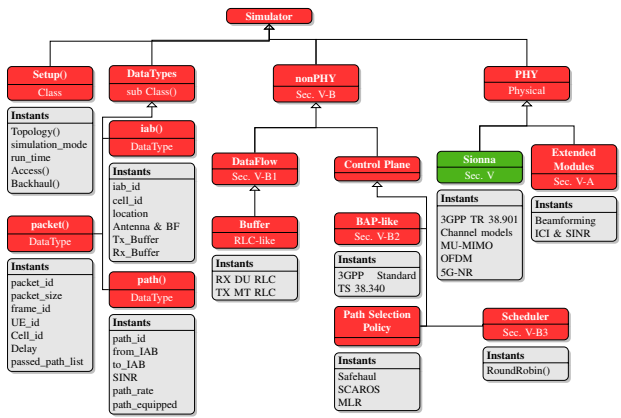
\subsection{System-level extensions to Sionna}
\label{sub:nonPHY_simulator}
As mentioned, Sionna is mainly a physical layer simulator. However, to get closer to \gls{iab} networks as specified in Rel. 17, we have extended Sionna by implementing a selection of system-level features. To such end, we introduced a discrete-event network simulator for modeling \gls{iab} networks. This system-level extension operates atop Sionna and provides basic functionality such as a \gls{mac}-level scheduler, layer-2 buffers, and data flow and path selection mechanisms. Our simulator, as depicted in Fig.~\ref{fig:simulator_design}, generates a variety of system-level KPIs such as latency, throughput, and packet drop rate. 

\subsubsection{Data Flow and buffer}
\label{sub:Dataflow}
3GPP has opted for a layer 2-relaying architecture for \gls{iab}-nodes where hop-by-hop \gls{rlc} channels are established. 
This enables retransmissions to take place just over the afflicted hops, thus preventing the need for traversing the whole route from the \gls{iab}-donor whenever a physical layer \gls{tb} is not decoded. Consequently, this design results in a more efficient recovery from transmission failures and reduces buffering at the communication endpoints~\cite{madapatha2020integrated}. To mimic this architecture, we have implemented RLC-like buffers at each base station. Specifically, each \gls{iab}-node features layer-2 buffers for both receiving and transmitting packets.
For instance, the data flow for an uplink packet is the following.
The \gls{ue} generates packets and sends a transmission request to the base station. Consequently, the scheduler allocates OFDM symbols for this transmission, which is eventually received and stored at the RX buffer of its \gls{du}. Next, the packet is placed into the TX buffer to be forwarded to the suitable next hop \gls{iab}-nodes. This procedure is repeated until the packet crosses all the wireless-backhaul hops and reaches the \gls{iab}-donor. Note that the packet can be dropped due to a violation of latency constraints or interference.

\subsubsection{Backhaul Adaptation Protocol}
\label{sub:bap}
To manage routing within the wireless-backhauled network, the 3GPP introduced the \gls{bap}, i.e., an adaptation layer above \gls{rlc} which is responsible for packet forwarding between the \gls{iab}-donor and the access \gls{iab}-nodes~\cite{3gpp_38_340}. Our simulator mimics this by associating each \gls{iab}-node to a unique \gls{bap} ID. Moreover, we append a \gls{bap} routing ID to each packet at its entry point in the \gls{ran} (i.e., the \gls{iab}-donor and the \glspl{ue} for DL and UL data, respectively).  Then, this identifier is used to discern the (possibly multiple) routes toward the packet's intended destination~\cite{3gpp_38_340}. The choice of the specific route is managed by \name{}.


	


\subsubsection{Scheduler}
\label{sub:scheduler}
Finally, we implement a \gls{mac}-level scheduler which operates in a \gls{tdma} mode. The scheduler periodically allocates the time resources to backhaul or access transmissions in a Round-Robin fashion. Specifically, each cell first estimates the number of OFDM symbols needed by each data flow by examining the corresponding buffer. Then, the subframe's OFDM symbols are equally allocated to the users. If a user requires fewer symbols to transmit its complete buffer, the excess symbols (the difference between the available slot length and the needed slot length) are dispersed to the other active users.

\section {Performance Evaluation}
\label{s:simulation_analysis}

\begin{figure}[t!]
    \centering
    \includegraphics[width=0.9\columnwidth]{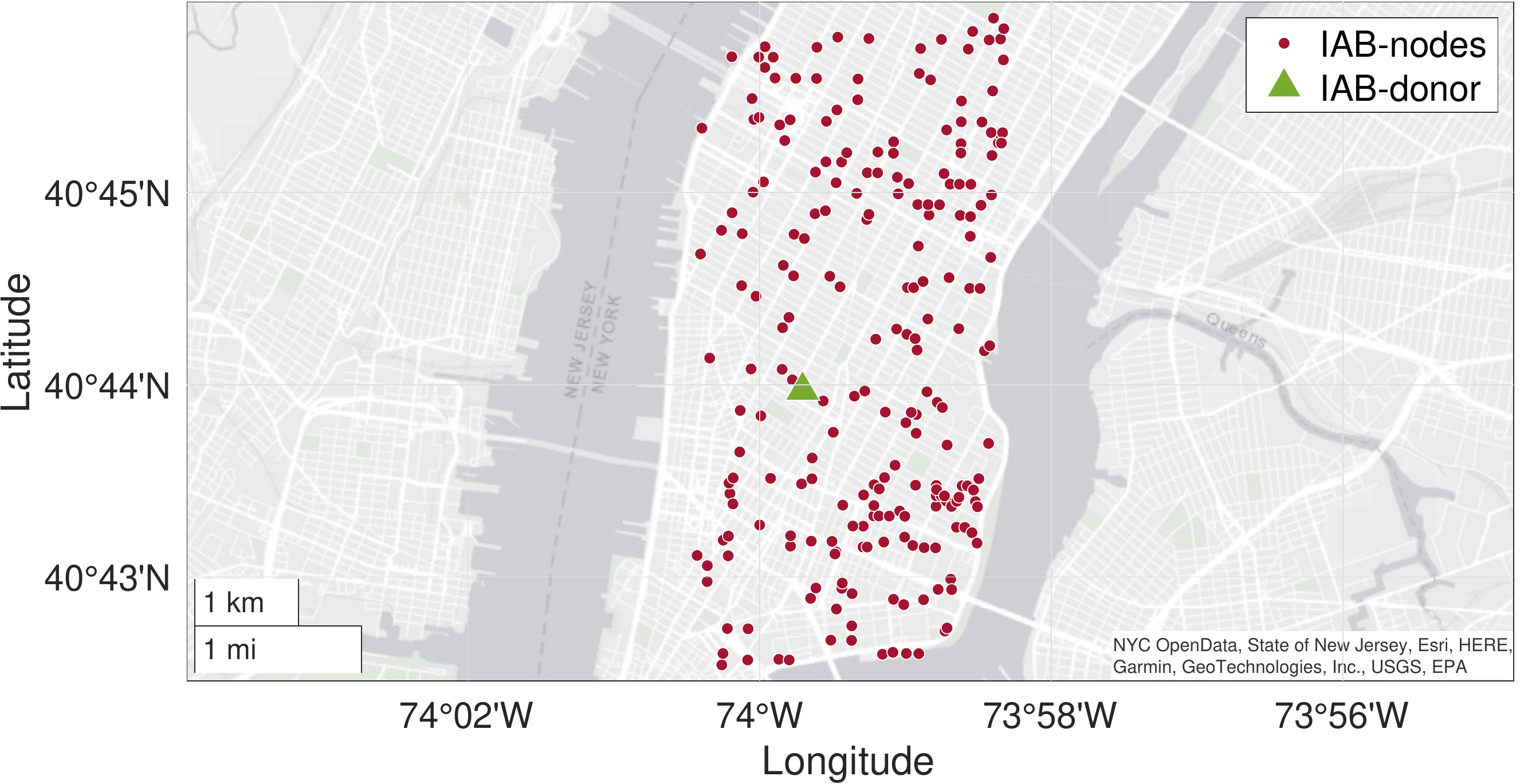}
      \caption{Locations of the 223 \node{} and \donor{} in Manhattan, NYC.}
      \label{fig:manhattan}
\end{figure}


\begin{table}[t!]
\caption{Simulation parameters.}
\label{Tab:parameters}
\centering
\footnotesize
\begin{tabular}{l|l}
    \toprule
    Parameter & Value \\ \midrule

    Carrier frequency and bandwidth & 28~GHz and 400 MHz \\
    IAB RF chains &   2 (1 access + 1 backhaul) \\
    Pathloss model & UMi-Street Canyon \cite{3gpp.38.901} \\
    Number of \node{} $N$ & $223$   \\
    Source rate & \{40, 80\} Mbps \\
    IAB Backhaul and access antenna array & 8H×8V and 4H×4V\\
    UE  antenna array & 4H×4V\\
    IAB and UE height & 15~m and 1.5~m \\
    IAB antenna gain & 33~dB \\
    Noise power & 10~dB \\
    Risk level $\alpha$ & 0.1\\
    Reliability weight factor $\eta$ & 1\\
\bottomrule
\end{tabular}
\end{table}

\begin{figure*}[t!]
\centering
    \subfloat[Average per-UE end-to-end latency]{
    \includegraphics[width=.3\linewidth,height=0.5\columnwidth]{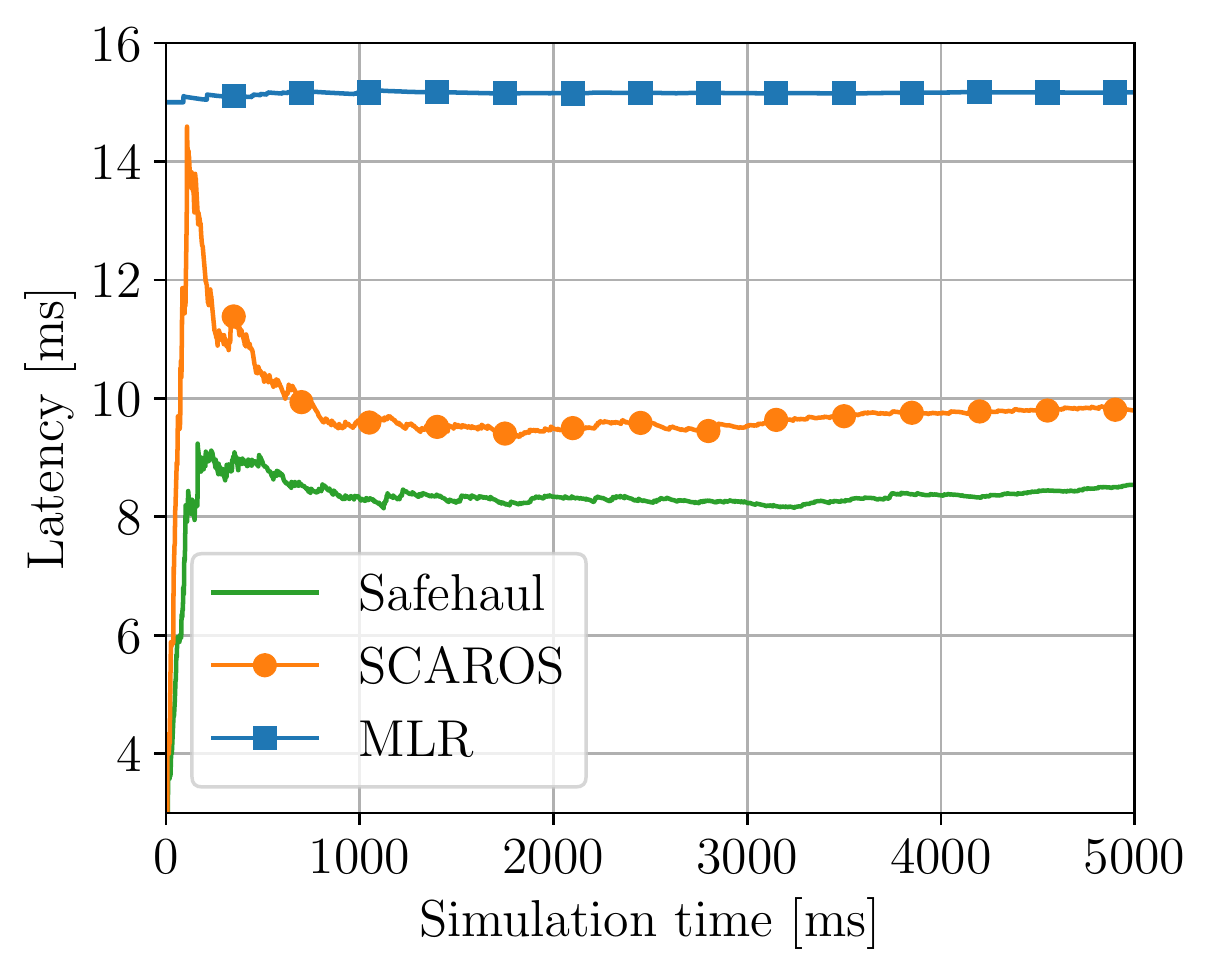}
    \label{fig:avgE2EDelay}
    }
    \subfloat[Average per-UE throughput]{
    \includegraphics[width=.3\linewidth,height=0.5\columnwidth]{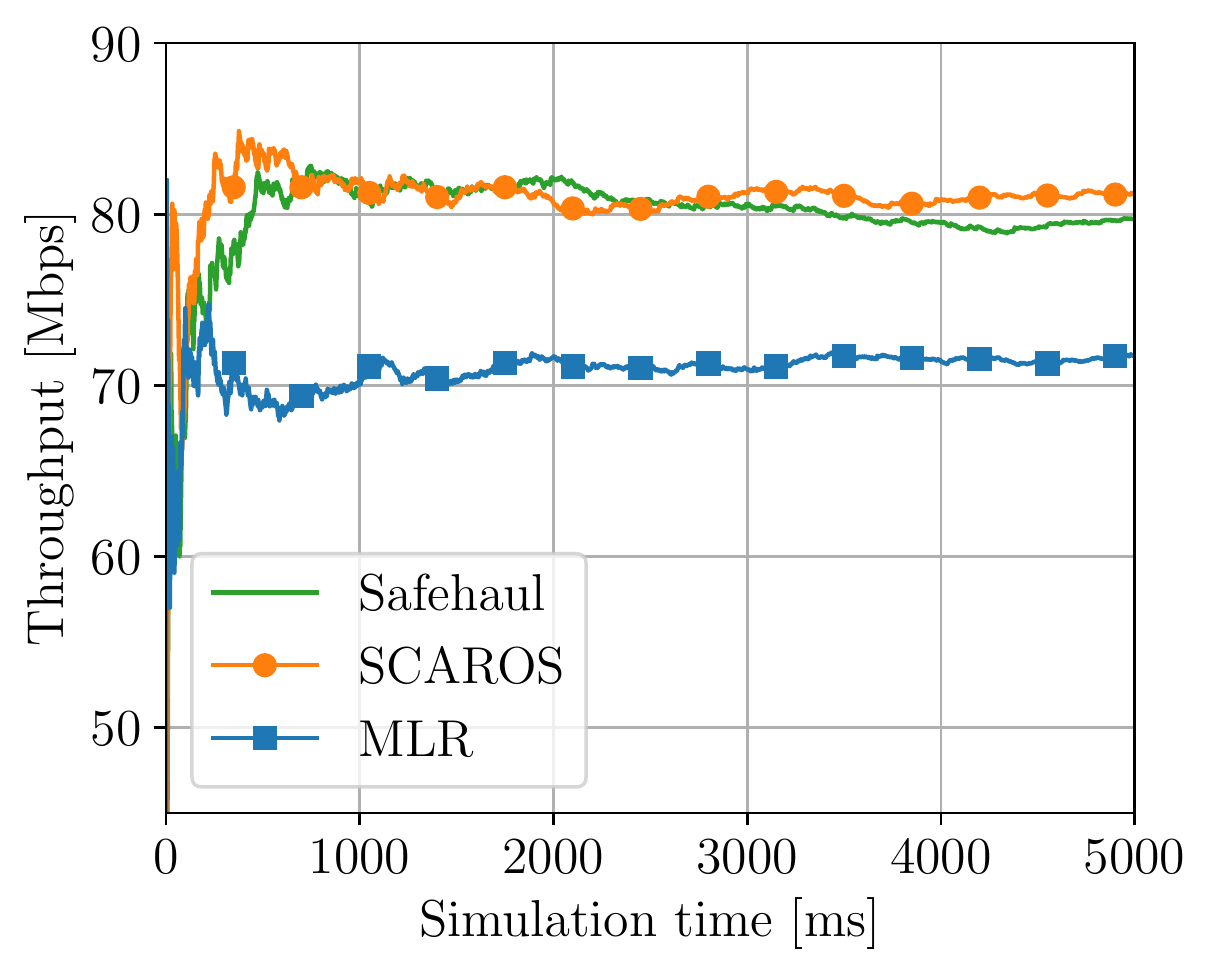}
    \label{fig:avgTput}
    }
    \subfloat[Average per-UE packet drop rate]{
    \includegraphics[width=.3\linewidth,height=0.5\columnwidth]{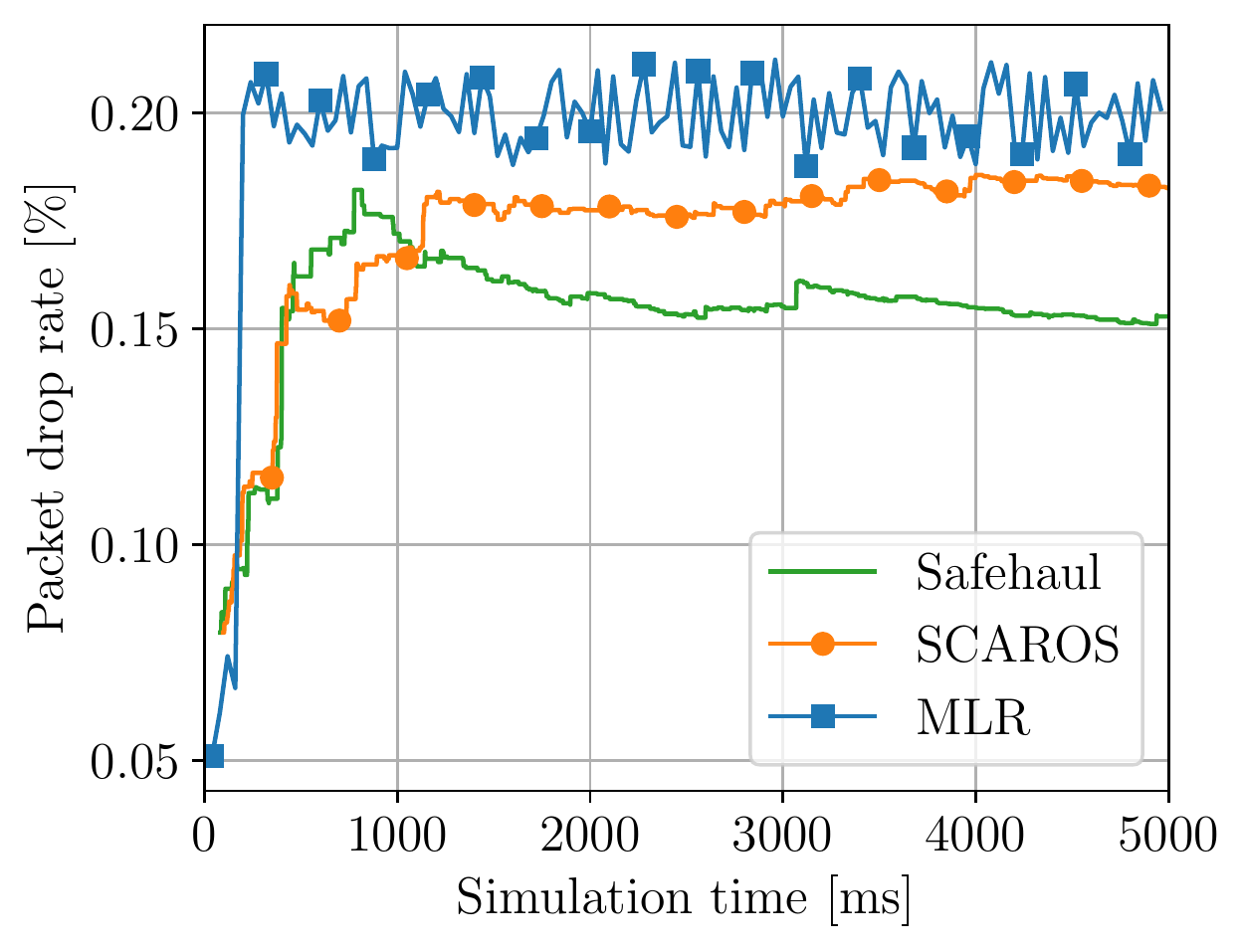}
    \label{fig:avgDropRate}
    }
  
   \caption{Average network performance for $100$ \glspl{ue} and $80$~Mbps per-UE source rate (Scenario 1).}
  \label{fig:avgNetPerfomance}
\end{figure*}
In our simulations, we consider a realistic cellular base station deployment in Manhattan, New York City\footnote{The locations correspond to the network of T-Mobile, as it has the largest deployment among the operators.}. Specifically, we collect the locations of $N=223$ 5G-NR base stations in an area of 15~$\text{Km}^2$ as depicted in Fig.~\ref{fig:manhattan}. The detailed simulation parameters are provided in Table \ref{Tab:parameters}. We used the channel model outlined by 3GPP in TR 38.901 \cite{3gpp.38.901}, which provides a statistical channel model for 0.5-100 GHz, and analyzed the "Urban Micro (UMi)-StreetCanyon" scenario.

\begin{figure*}
\centering
    \subfloat[Per-UE end-to-end Latency]{
    \includegraphics[width=.3\linewidth,height=0.5\columnwidth]{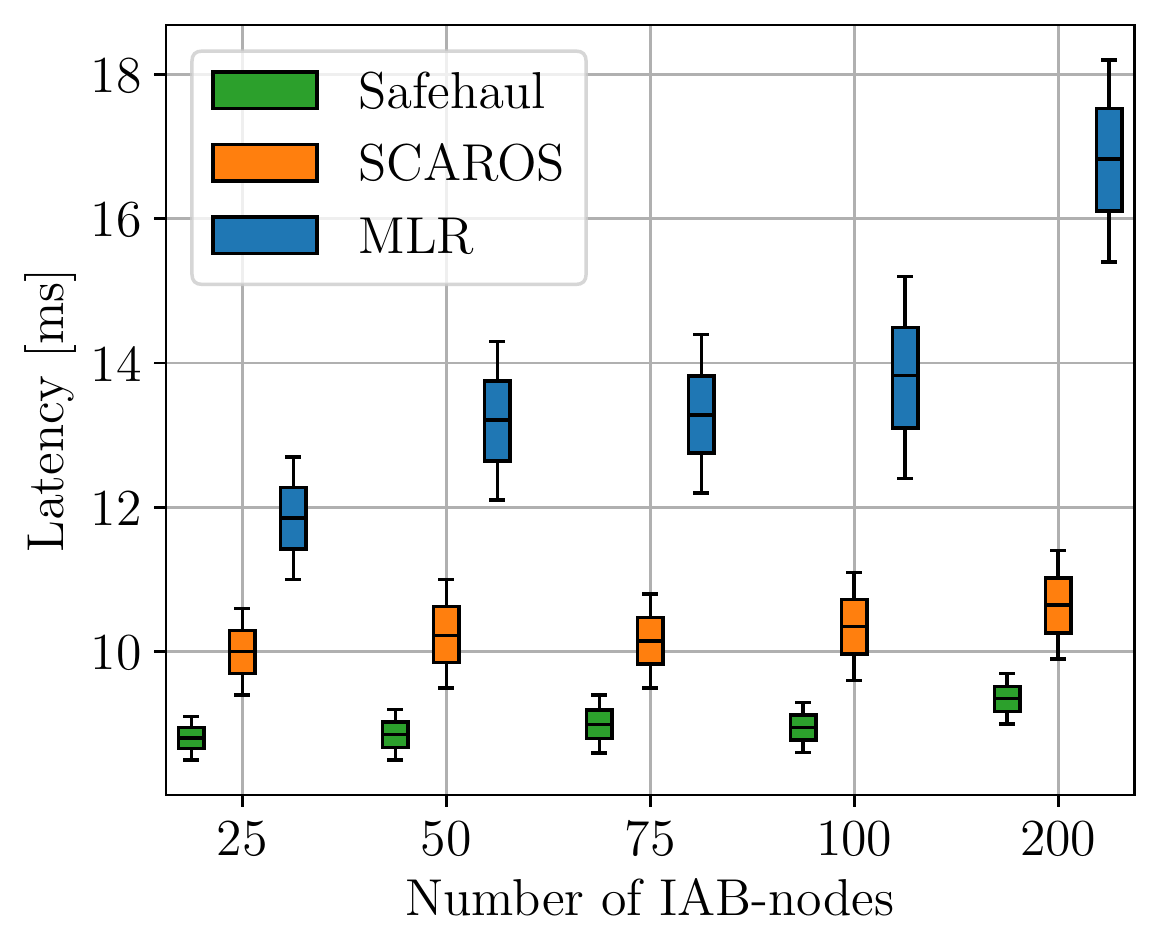}
    \label{fig:avgE2EDelay_s2}
    }
    \subfloat[Per-UE throughput]{
    \includegraphics[width=.3\linewidth,height=0.5\columnwidth]{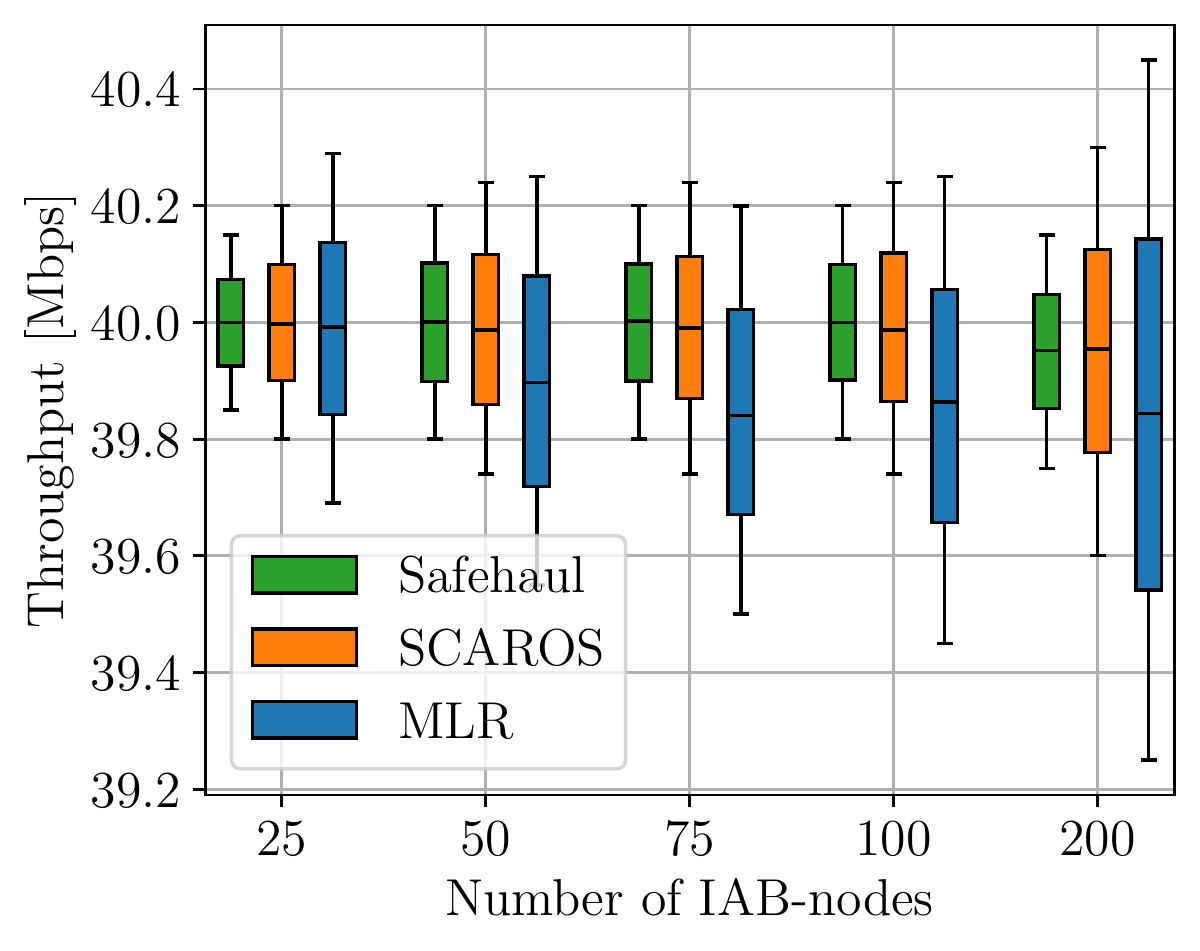}
    \label{fig:avgTput_s2}
    }
    \subfloat[Per-UE packet drop rate]{
    \includegraphics[width=.3\linewidth,height=0.5\columnwidth]{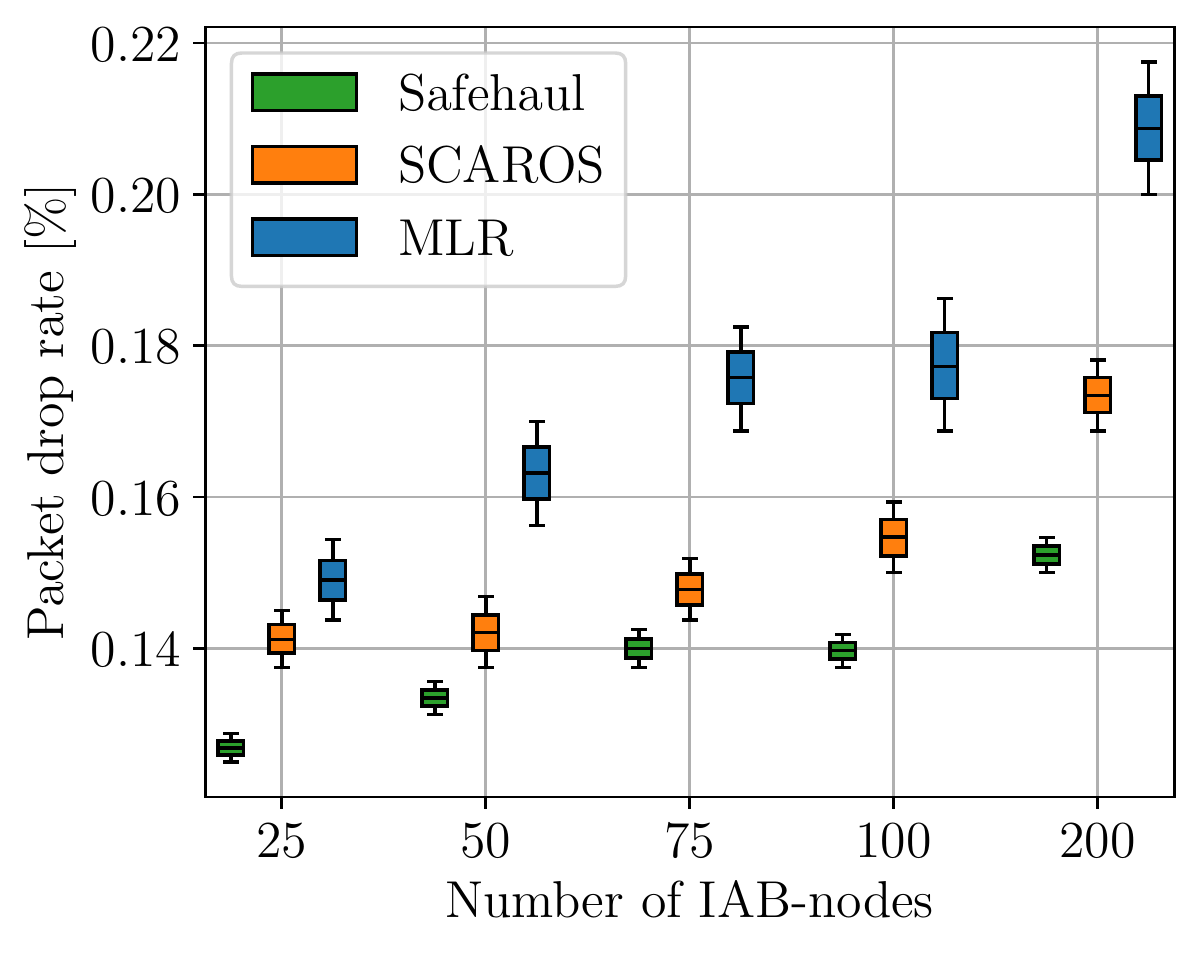}
    \label{fig:avgDropRate_s2}
    }
  
   \caption{Network performance for $\{25, 50, 75, 100, 200\}$ \gls{iab}-nodes and $40$ Mbps per-UE source rate (Scenario 2).}
  \label{fig:avgNetPerfomance_s2}
\end{figure*}

\begin{figure*}
\centering
    \subfloat[Average per-UE end-to-end Latency]{
    \includegraphics[width=.3\linewidth,height=0.5\columnwidth]{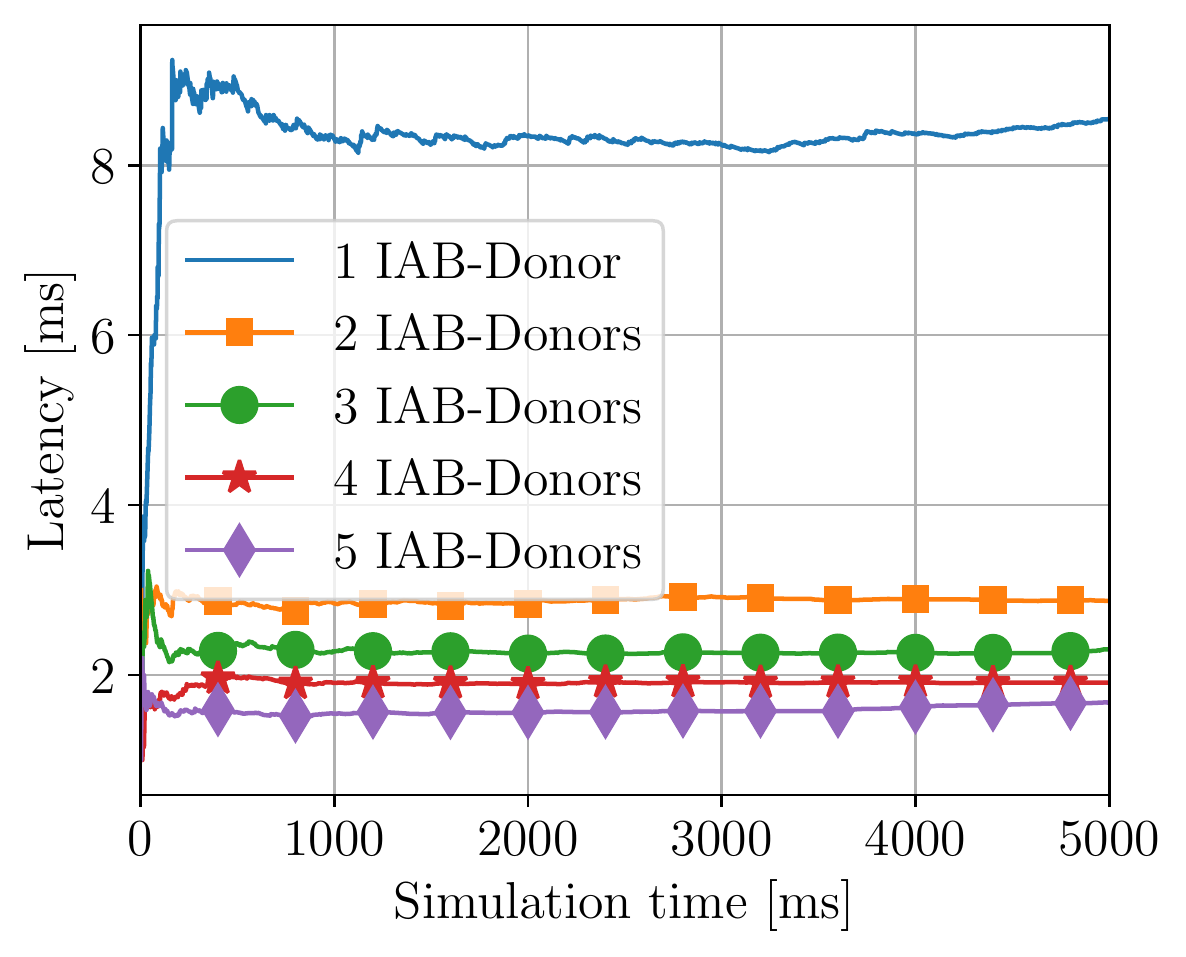}
    \label{fig:avgE2EDelay_s3}
    }
    \subfloat[Average per-UE throughput]{
    \includegraphics[width=.3\linewidth,height=0.5\columnwidth]{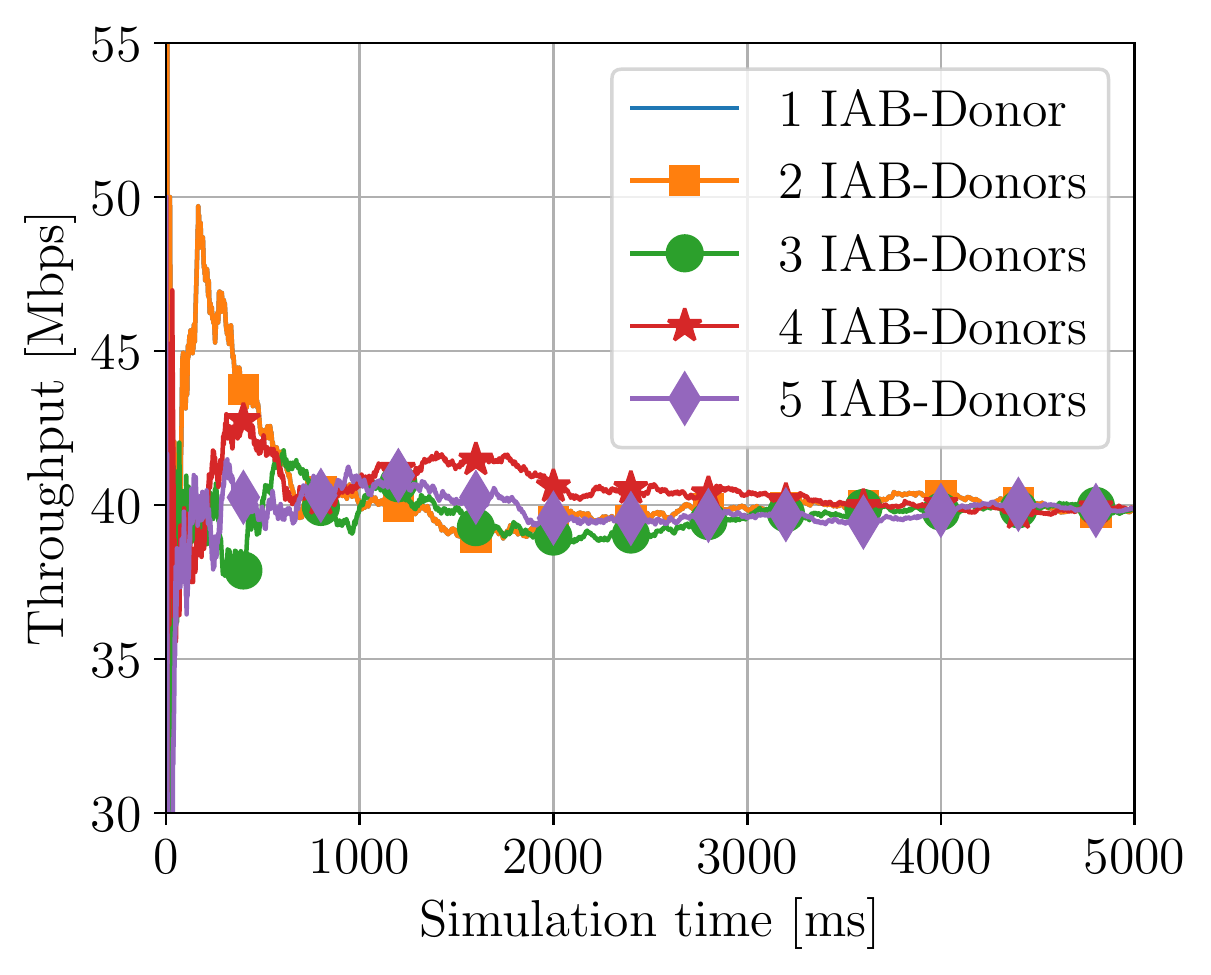}
    \label{fig:avgTput_s3}
    }
    \subfloat[Average per-UE packet drop rate]{
    \includegraphics[width=.3\linewidth,height=0.5\columnwidth]{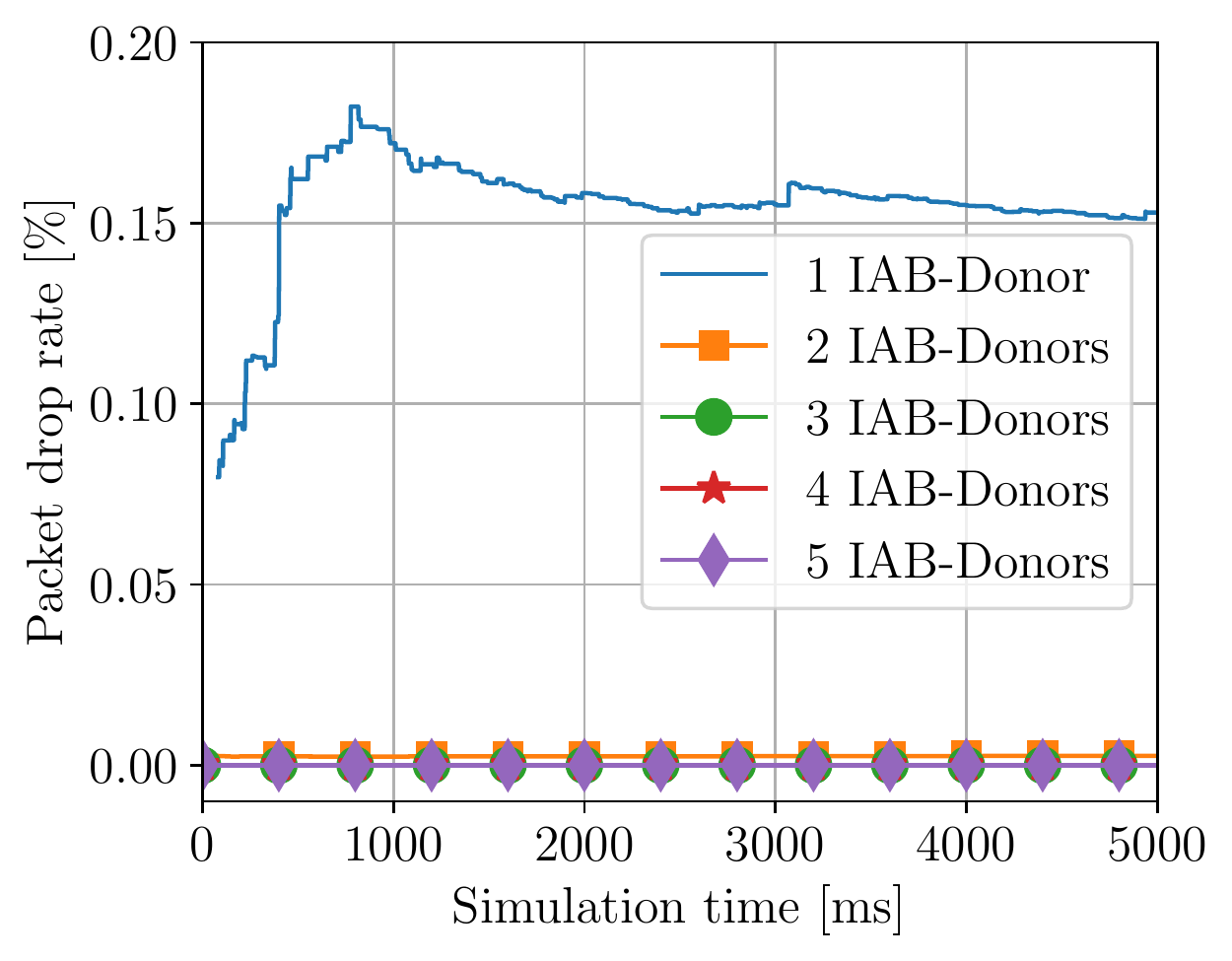}
    \label{fig:avgDropRate_s3}
    }
  
   \caption{Network performance for $100$ \glspl{ue} and $40$ Mbps per-UE source rate, versus the number of \gls{iab}-donors (Scenario 3).}
  \label{fig:avgNetPerfomance_s3}
\end{figure*}
\textbf{Benchmarks.} To provide better insights on the performance of \name{}, we replicate two approaches from the state of the art: $(i)$ \gls{scaros}, a learning-based approach that minimizes the average latency in the network~\cite{ortiz2019scaros}, and $(ii)$ \gls{mlr}, a greedy approach aiming to maximize throughput by selecting the links with the highest data rate.
Our evaluation consists of four scenarios studying the convergence of the algorithms to a steady state, the number of IAB-nodes, the number of IAB-donors, and the impact of risk aversion. When demonstrating the results, we show the average throughput, latency, and packet drop rate per UE. Furthermore, we show the statistical variance of the obtained results using candlesticks which include the max, min, mean, and 10 and 90 percentiles of the achieved performance.


\subsection{Scenario 1: Average Network Performance}
\label{sub:netPerformance}
Analyzing the performance of the algorithms as a function of time is crucial to determine the convergence speed of the learning-based techniques, i.e.,  \name{} and \gls{scaros}. Hence, in Fig.~\ref{fig:avgNetPerfomance} we show the average network performance over time for three metrics: latency, throughput, and packet drop rate.

In Fig.~\ref{fig:avgE2EDelay}, we can observe that \name{} rapidly converges to an average latency of approximately $8.6$ ms which is $12.2$\% and $43.4$\% lower than the latency of \gls{scaros} and \gls{mlr}, respectively. The high performance of \name{} stems from the joint minimization of the average latency and the expected value of its tail loss, which results in avoiding risky situations where latency goes beyond $T_\mathrm{max}$.
This is not the case for \gls{scaros} where we observe a high peak in the latency before convergence, i.e., in between zero and 1000~ms. \textit{It is exactly the avoidance of such transients in \name{} that leads to higher reliability in the system.} The reliability offered by \name{} allows operators to deploy self-backhauling in an online fashion and without disrupting the network operation. Moreover, it protects the networks from the transients that may arise from changes in the network topology.
The performance of \gls{mlr} is constant throughout the simulation, as it is not designed as an adaptive algorithm.

Figure \ref{fig:avgTput} shows that the risk-aversion capabilities of \name{} have no negative impact on the average throughput in the network. 
The performance of \name{} is comparable to that of \gls{scaros}, approximately $79.3$ Mbps, and $11.7$\% larger than the performance of \gls{mlr}.

The performance shown in Figure \ref{fig:avgDropRate} is consistent with the behaviour observed in Figure \ref{fig:avgE2EDelay}. As \name{} additionally minimizes the $\alpha$-worst latency, it achieves the lowest packet drop rate, compared to the reference schemes, namely, $16.6$\% and $25.0$\% lower than \gls{scaros} and \gls{mlr}, respectively.


\subsection{Scenario 2: Impact of Network Size}
\label{sub:netSize}
In Fig.~\ref{fig:avgNetPerfomance_s2} we evaluate the reliability of the three considered approaches for different network sizes. Specifically, we vary the number of \node{} starting from 25 up to 100.  At the same time, we increase the load in the network by increasing the number of \gls{ue}s.
From the figures, we can clearly see that \name{} consistently achieves a lower variation compared to the reference schemes. This verifies that \name{} achieves the intended optimization goal, i.e.,  the joint minimization of the average performance and the worst-case losses.

Fig.~\ref{fig:avgE2EDelay_s2} shows that \name{} is able to maintain an almost constant latency as the number of \node{} increases. Specifically, the variation of latency with \name{} is  56.1\% and 71.4\% less than \gls{scaros} and \gls{mlr} , respectively. Furthermore, \name{} achieves 11.1\% and 43.2\% lower latency compared to \gls{scaros} and \gls{mlr}. \gls{mlr}'s high variance is due to a lack of adaptation capabilities, hence its latency variance is governed by the network's underlying random processes.

As shown in Fig.~\ref{fig:avgTput_s2}, the average throughput of the learning-based approaches \name{} and \gls{scaros} remains constant for the different values of network size. However, the lowest variation in the  throughput is achieved by \name{}, i.e., only 0.40 compared to 0.51 and 0.79 in the benchmark schemes. Such behaviour corroborates \name{}'s reliability capabilities. 

The packet drop rate for different number of \gls{iab}-nodes is shown in Fig.~\ref{fig:avgDropRate_s2}. \name{} not only consistently outperforms the reference schemes, but also with the minimum variation in the results (by at least 39.1\% compared to benchmarks). Considering the largest network size and load, i.e., 200 \node{} and 400 \glspl{ue}, \name{} achieves 11.2\% and 24.9\% lower packet drop rate compared to \gls{scaros} and \gls{mlr}, respectively.

\subsection{Scenario 3: Impact of number of IAB-donors}
\label{sec:multiDonors}
Although the benchmark schemes do not support multi-IAB-donors, \name{} is designed to accommodate such scenarios. In Fig.~\ref{fig:avgNetPerfomance_s3}, we investigate the impact of the number of IAB-donors on \name{}. Specifically, the network load is constant in this scenario, i.e., the number of \glspl{ue} is fixed. 

We observe in Fig.~\ref{fig:avgE2EDelay_s3} that the highest latency is experienced when only one IAB-donor is present in the network. This stems from the tributary effect of self-backhauling where the traffic flows towards a central entity which itself can become a bottleneck. As the number of IAB-donor increases, the traffic flow is more evenly distributed, resulting in lower latency. Specifically, from an average latency of 8.2 ms for $D=1$, to an average latency of 1.7 ms when $D=5$. 
As mentioned, since the load is constant in this scenario, the average throughput remains also constant for all different numbers of IAB-donors, see Fig.~\ref{fig:avgTput_s3}. However we should highlight that \name{}'s  learning speed is maintained for the different values of $D$.
This is an important design feature of \name{} because having more \donor{} means that the number of paths a \node{} has to the core network increases exponentially. From a learning perspective, such increment implies a larger action set and a lower learning speed. \name{} avoids this problem by learning the average latency based on the estimates of its neighbors and not on the complete paths to the \donor{}.
Finally, Fig. \ref{fig:avgDropRate_s3} shows that increasing the number of \donors~significantly reduces the packet drops, which also stems from a better distribution of traffic flows in the network as observed in Fig.~\ref{fig:avgE2EDelay_s3}. 

\begin{figure}[ t!]
    \centering
    \includegraphics[width=0.6\columnwidth]{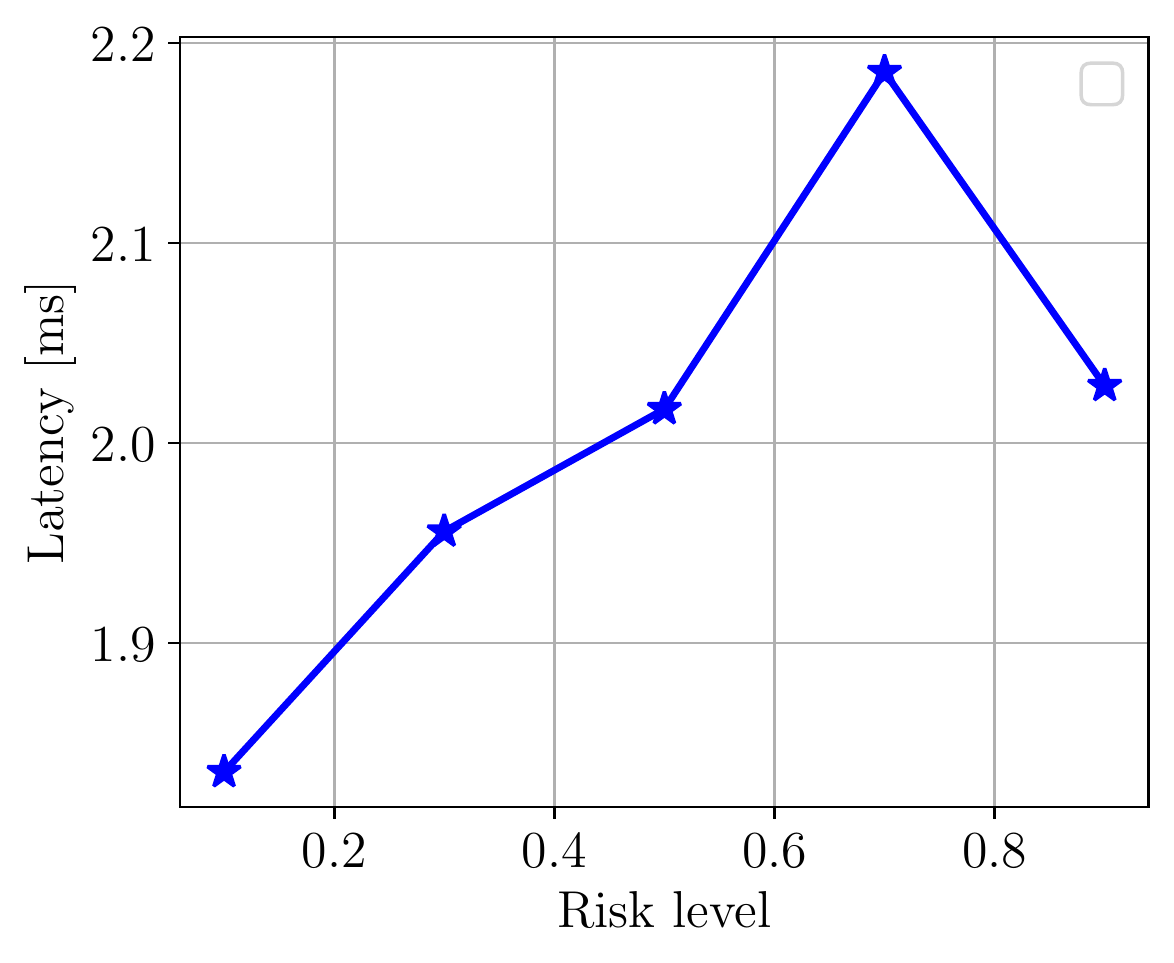}
      \caption{Average latency for 100 UEs and 20 Mbps per-UE source rate, versus the risk level $\alpha$ (Scenario 4)}
      \label{fig:riskParam}
\end{figure}

\subsection{Scenario 4: Impact of risk parameter $\alpha$}
The definition of losses in the tail of the latency distribution is controlled by the risk level parameter $\alpha$. Its impact on the average latency is shown in Fig. \ref{fig:riskParam}, where an increasing behaviour is observed for $\alpha\leq 0.7$. The lowest latency is achieved for $\alpha=0.1$, which corresponds to the most risk-averse, and therefore the most reliable, case out of all the considered ones. As $\alpha$ grows, the performance of \name{} tends to that of the risk-neutral case.




\section {Related work}
\label{s:related}




Self-backhauling wireless networks have been studied in different contexts. Ranging from the so-called \glspl{hetnet} and \gls{iab} 5G \gls{nr} systems, to \glspl{cran}, each has considered a different set of premises and optimization goals. In this section, we review the related work in the context of basic assumptions and their optimization goals. Furthermore, we shed light on some of the common but perhaps unrealistic assumptions which we refrain from in this article. 

\textbf{Ideal backhaul links.} Numerous works assume either an \textit{infinite or fixed capacity backhaul link}. This is often motivated due to the presence of a wired fiber link between the \glspl{sbs} and the \gls{mbs}~\cite{pan2017joint, huang2015joint, nguyen2020nonsmooth, rasekh2015interference}. Indeed, most of these works consider a scenario where a centralized \gls{bbu} is connected to several \glspl{rrh}, i.e., radios which lack signal processing capabilities~\cite{pan2017joint, huang2015joint, nguyen2020nonsmooth}. In particular, the authors of~\cite{nguyen2020nonsmooth} consider an even more complex scenario referred to as F-RAN, i.e., a C-RAN where RRHs feature caching and signal processing capabilities. However, in an \gls{iab} context it is fundamental to consider \textit{limited-rate, time-varying backhaul channels} and to study the impact of such limitations on the performance of the \gls{ran}.

\textbf{Constrained topologies.} It is often assumed that self-backhauled networks have a \textit{specific topology}. This assumption usually simplifies the problem and makes it tractable and/or solvable in polynomial time. For instance, the authors of~\cite{kwon2019joint, pizzo2017optimal, lei2020deep} assume a single-hop network where each \gls{sbs} is directly connected to the \gls{mbs}. In~\cite{kulkarni2018max}, a $k$-ring deployment is considered, i.e., a topology where a single \gls{iab}-donor provides backhaul connectivity to $k$ rings of \gls{iab}-nodes. Even though this topology can be used to model networks with arbitrary depth, it maintains a symmetric load for each node, an assumption which generally does not hold in real networks.
In fact, the 3GPP does not impose any limits on the number of \gls{iab}-nodes which can be connected to a given \gls{iab}-donor, nor does it set an upper bound on the number of wireless hops from the latter to other wireless-backhauled base stations~\cite{3gpp_38_874}. Accordingly, in our problem formulation we consider \gls{iab} networks with an \textit{arbitrary number of nodes and an arbitrary maximum wireless hops} between \glspl{mbs} and \glspl{sbs}. 

\textbf{Simplistic traffic models.} Some works assume either a \textit{full buffer traffic model and/or impose flow conservation constraints}. In particular, the authors of~\cite{yuan2018optimal, rasekh2015interference} consider systems where the capacity of each link can  always be fully exploited thanks to the presence of \textit{infinite data to transmit at each node}. However, in actual \gls{iab} deployments the presence of packets at the \glspl{mbs} and \glspl{sbs} is conditioned on the \textit{status of their \gls{rlc} buffers and, in turn, on the previous scheduling decisions}. Moreover, \textit{packets can actually be buffered at the intermediate nodes}, thus preventing the need for transmitting a given packet in consecutive time instants along the whole route from the \gls{iab}-donor to the \glspl{ue} (or vice versa). 

\textbf{Optimization goals.} The works in the literature focus on different optimization goals. Therefore, they prioritize different network metrics. For instance, the authors of~\cite{hur2013millimeter} aim to optimize the beam alignment between \glspl{mbs} and \glspl{sbs}. Instead, the work of~\cite{alizadeh2019load} aims to compute the optimal user-to-base-station association. However, they neglect backhaul associations and focus on the access only. In~\cite{kwon2019joint, alizadeh2019load, zhu2016qos, yuan2018optimal} the objective function is a function of the users data-rate. In particular, the authors of~\cite{yuan2018optimal} optimize the max-min user throughput, arguing that such a metric better captures the performance of the bottleneck links. In~\cite{vu2018path}, the average rate of each link is maximized under bounded delay constraint. In our work, we focus on reliability by minimizing not only the average end-to-end delay, but also the expected value of the worst-case performance.
The work closest to this article is SCAROS~\cite{ortiz2019scaros},  a learning-based latency-aware scheme for resource allocation and path selection in self-backhauled networks. Assuming a single IAB-donor, the authors study arbitrary multi-tier \gls{iab} networks considering the impact of interference and network dynamics. In contrast to this work, we aim at enhancing the reliability of the IAB-network by jointly minimizing the average end-to-end delay and its expected tail loss. Moreover, considering realistic deployments, our proposed \name{} supports networks with an arbitrary number of IAB-donors.

\section{Conclusion}
\label{s:conclusion}
In this work, we proposed the first reliability-focused scheduling and path selection algorithm for \gls{iab} mmWave networks. Via extensive simulations, we illustrated that our \gls{rl}-based solution can cope with the network dynamics including channel, interference, and load.  Furthermore, we demonstrated that \name{} not only  exhibits highly reliable performance in the presence of the above-mentioned network dynamics but also, it outperforms the benchmark schemes in terms of throughput, latency and packet-drop rate. The reliability of \name{} stems from the joint minimization of the average latency and the expected value of its tail losses. The latter is achieved by leveraging \gls{cvar} as a risk metric.


Reliability is a highly under-explored topic that definitely deserves more investigation. Some interesting research directions are the maximization of reliability under the assumption of statistical system knowledge, or the evaluation of the network's reliability when the functionality of the \gls{bap} layer is compromised.
Furthermore, our system-level extension to Sionna can be further developed to support an arbitrary number of RF chains and in-band backhauling, allowing more extensive investigation of IAB protocols and architecture.

\section{Acknowledgement}
This research was partly funded by the Deutsche Forschungsgemeinschaft within the mm-Cell project and the Collaborative Research Center 1053 MAKI, by the LOEWE initiative (Hesse, Germany) within the emergenCITY center, the Bundesministerium f\"ur Bildung und Forschung through the Open6GHub project and by the European Commission through Grant No. 861222 (H2020 ITN MINTS project).

\bibliographystyle{IEEEtran}

\bibliography{IEEEabrv, bibl}

\end{document}